\documentclass[12pt]{article}

\usepackage{amsmath,amsfonts,amsthm,subfloat,algorithm,url,float,epsf,psfrag}

\usepackage[pdftex]{color,graphicx}
\usepackage[colorlinks,citecolor=blue,urlcolor=blue]{hyperref}

\usepackage[round]{natbib}

\usepackage{setspace}
\usepackage{blindtext}
 \usepackage{natbib}
\usepackage[small,bf]{caption}
\usepackage[utf8]{inputenc}

\usepackage{mathrsfs}
\usepackage{amsfonts}
\usepackage{enumerate}
\usepackage{latexsym}
\usepackage{multirow}

\floatname{algorithm}{Algorithm}

\newcommand{\ML}{{\mathcal L}}

\newtheorem{thm}{Theorem} 
\newtheorem{lem}{Lemma}

\newtheorem{mydef}{Definition}
\newtheorem{rem}{Remark}

\newcommand{\pr}{\mathbb{P}}
\newcommand{\E}{\mathbb{E}}

\newcommand{\B}{\boldsymbol}
\newcommand{\M}{\mathbf}
\newcommand{\X}{\mathcal{X}}
\newcommand{\C}{\mathcal{C}}

\newcommand{\eps}{\epsilon}

\newcommand{\bt}{\pmb \theta}
\newcommand{\imply}{\Longrightarrow}
\newcommand{\half}{\mbox{$\frac12$}}

\usepackage{graphicx}
\usepackage{amsmath}
\makeatletter
\newcommand{\distas}[1]{\mathbin{\overset{#1}{\kern\z@\sim}}}%
\newsavebox{\mybox}\newsavebox{\mysim}
\newcommand{\distras}[1]{%
  \savebox{\mybox}{\hbox{\kern3pt$\scriptstyle#1$\kern3pt}}%
  \savebox{\mysim}{\hbox{$\sim$}}%
  \mathbin{\overset{#1}{\kern\z@\resizebox{\wd\mybox}{\ht\mysim}{$\sim$}}}%
}
\makeatother

\DeclareMathOperator*{\argmin}{arg\,min}
\DeclareMathOperator*{\argmax}{arg\,max}
\DeclareMathOperator*{\mini}{minimize}

\usepackage{xr}
\usepackage{enumerate}
\newcommand{\blind}{1}

\begin{document}

\def\spacingset#1{\renewcommand{\baselinestretch}%
{#1}\small\normalsize} \spacingset{1}


\if1\blind
{
  \title{\LARGE \bf A Computational Framework for Multivariate Convex Regression and its Variants}
  \author{Rahul Mazumder\thanks{Supported by ONR grant N00014-15-1-2342 and an interface grant from the Betty-Moore Sloan Foundation. E-mail: rahulmaz@mit.edu}\hspace{.2cm}\\
    \small{MIT Sloan School of Management and Operations Research Center,}\\
     \small{Massachusetts Institute of Technology}    \\ 
  Arkopal Choudhury\thanks{E-mail: arkopal@live.unc.edu} \hspace{.2cm}\\
    \small{Department of Biostatistics, University of North Carolina at Chapel Hill} \\
    Garud Iyengar\thanks{Supported by NSF: DMS-1016571, CMMI-1235023, ONR: N000140310514; E-mail: garud@ieor.columbia.edu} \hspace{.2cm}\\
    \small{IEOR Department, Columbia University} \\
    Bodhisattva Sen\thanks{Supported by NSF CAREER Grant DMS-1150435; e-mail: bodhi@stat.columbia.edu} \hspace{.2cm}\\
    \small{Department of Statistics, Columbia University}     
    }
  \maketitle
} \fi

\if0\blind
{
  \bigskip
  \bigskip
  \bigskip
  \begin{center}
    {\LARGE\bf A Computational Framework for Multivariate Convex Regression and its Variants}
\end{center}
  \medskip
} \fi

\bigskip
\begin{abstract}
We study the nonparametric least squares estimator (LSE) of a multivariate convex regression function. The LSE, given as the solution to a quadratic program with $O(n^2)$ linear constraints ($n$ being the sample size), is difficult to compute for large problems. Exploiting problem specific structure, we propose a scalable algorithmic framework based on the augmented Lagrangian method to compute the LSE. We develop a novel approach to obtain smooth convex approximations to the fitted (piecewise affine) convex LSE and provide formal bounds on the quality of approximation. When the number of samples is not too large compared to the dimension of the predictor, we propose a regularization scheme ---  Lipschitz convex regression --- where we constrain the norm of the subgradients, and study the rates of convergence of the obtained LSE. Our algorithmic framework is simple and flexible and can be easily adapted to handle variants: estimation of a non-decreasing/non-increasing convex/concave (with or without a Lipschitz bound) function. We perform numerical studies illustrating the scalability of the proposed algorithm.
\end{abstract}

\noindent





\spacingset{1.2} 

\section{Introduction}
Consider the task of fitting a {\it multivariate convex} function to observations $\{(\M X_i, Y_i)\}_{i=1}^n$ where the covariates $\M X_i \in \Re^d$, $d \ge 1$, and the response $Y_i \in \Re$. This problem has been recently considered by several authors; see e.g., \cite{SS11} and \cite{LG12} and the references therein. Convex (concave) regression problems are common in economics, operations research and reinforcement
learning. In economics, production functions and utility function preferences are usually known to be concave (see~\cite{AllonEtAl07} and \cite{V84}) whereas consumer preferences (see~\cite{MP68}) are often assumed convex. In operations research and reinforcement learning, value functions for stochastic optimization problems can be convex (see~\cite{ShapiroEtAl09}). 

Probably the most natural estimator of the convex regression function in this setup is the least squares estimator (LSE) $\hat \phi_n$, defined as a minimizer of the empirical $L_2$-norm, i.e.,  
\begin{equation}\label{eq:LSE}
\hat \phi_n \in \argmin \sum_{i=1}^n (Y_i - \psi(\M{X}_i))^2,
\end{equation} 
where the minimum is taken over all convex functions $\psi: \Re^d \rightarrow \Re$. See \citet{SS11} and \citet{LG12} for the characterization, computation and consistency of the LSE.
An appealing feature of the convex LSE over most other nonparametric function estimation methods is that no tuning parameters  (e.g., smoothing bandwidths as in kernel regression) need to be specified in the estimation procedure.

The seemingly infinite dimensional optimization problem in \eqref{eq:LSE} can indeed be reduced to a finite dimensional one as  described below. Letting $\hat \theta_i = \hat\phi_n(\M{X}_i)$, for $ i = 1,\ldots,n$, we can compute $\hat{\B \theta} = (\hat \theta_1, \ldots, \hat \theta_{n})$ by solving the following Quadratic Program (QP):
\begin{equation}\label{eq:CvxLSE}
\begin{aligned}
\mini_{\B\xi_1, \ldots, \B\xi_n;  \B\theta} & \;\;\;\; \frac{1}{2}\| \M{Y} -  \B{\theta} \|_2^2\\
s.t. \;\;\; &\;\;\;  \theta_{j} + \langle \B{\Delta}_{ij}, \B{\xi}_{j} \rangle \leq \theta_{i}; \;\;\; i \ne j \in \{1, \ldots, n\}, 
\end{aligned}
\end{equation}
where $\B\Delta_{ij} := \M{X}_{i} - \M{X}_j \in \Re^{d}$, $\M{Y} = (Y_1,\ldots, Y_n)^\top$, $\B \xi_i \in \Re^d$, $\B\theta=(\theta_1, \ldots, \theta_{n})^\top  \in \Re^{n}$, $\langle \cdot , \cdot\rangle$ denotes the usual inner product  and $\|\cdot \|_2$ denotes the standard Euclidean norm. 
In words, Problem~\eqref{eq:CvxLSE} estimates the function values at the points $\M{X}_{i}$'s and also delivers estimates of subgradients $\B\xi_{i}$'s of the function at the $n$ points $\M{X}_{i}$'s. A natural way to extend $\hat \phi_n$ to a convex function defined on the whole of $\Re^d$ is to use the following rule:
\begin{equation}\label{eq:MaxLSE}
\hat{\phi}_n(\M{x}) = \max_{j=1,\ldots,n} \left\{\hat \theta_j + \langle \M{x} - \M{X}_j, \hat{\B \xi}_j \rangle \right\}.
\end{equation} 
\citet{SS11} used off-the-shelf interior point solvers (e.g., {\texttt {cvx}}, {\texttt {MOSEK}}, etc.) to solve the optimization problem \eqref{eq:CvxLSE}. However, off-the-shelf interior point solvers are not scalable and quickly become prohibitively expensive for sample size $n \ge 300$ due to the presence of $O(n^2)$ linear constraints. 
This motivates the present study where we propose scalable first order methods based on modern convex optimization techniques and investigate the statistical and computational issues of variants of convex regression. 
During the course of this work, we became aware of the conference paper by~\cite{Aybat14} which studies algorithms for an approximation of Problem~\eqref{eq:CvxLSE} by adding a ridge regularization on the subgradients $\B\xi_{j}$'s -- the authors report computational savings over interior point methods for problem sizes up to $n = 1600$. Our algorithmic approach in this paper, however, is different and we demonstrate scalability to problems of larger size. 


We summarize our main contributions below.


{\bf {Algorithmic Framework:}}	We propose, in Section~\ref{sec:AlgoCvxReg}, an algorithmic framework based on the augmented Lagrangian method~\citep{bertsekas-99,Boyd} and first order optimization techniques~\citep{nesterov2004introductorynew}  to solve the Problem~\eqref{eq:CvxLSE}, which is a quadratic program (QP) with $O(n^2)$ constraints in $O(nd)$ variables. 
	Our algorithmic framework is scalable, yet general enough to accommodate variants of the problem (see the discussion below). 
The convergence properties of the algorithm are subtle, and are discussed. 
An important aspect of our algorithm is that it heavily exploits problem structure to enhance its computational scalability. This enables us to compute the convex LSE for problem sizes much beyond the capacity of off-the-shelf interior point methods. Our approach delivers moderate accuracy solutions for $n\approx 5000$ within a few minutes; and scales quite easily for problems with $n \approx 10000$ or more. When $d \ll n$, the computational cost of our algorithm is dominated by $O(n^2)$, stemming from matrix-vector multiplications --- the cost is indeed reasonable since Problem~\eqref{eq:CvxLSE} has $O(n^2)$ linear constraints. Note that solving Problem~\eqref{eq:CvxLSE} with off-the-shelf interior point methods has a complexity of $O(n^3d^3)$; see e.g.,~\citet{BV2004}.

	
{\bf{Smooth Estimators:}} A characteristic feature of the LSE $\hat \phi_n$, as defined in~\eqref{eq:MaxLSE}, is that it is piecewise affine and hence non-smooth.
This may be perceived as a possible drawback of the LSE, since in various statistical applications, smooth estimators are desired; see e.g.,~\cite{BD07},~\cite{AguileraEtAl11},~\cite{Du13} and the references therein.
	Most of the existing work on function estimation under both smoothness and shape constraints are for one dimensional functions --- estimating a smooth multivariate convex function is a challenging statistical problem.
We present in Section~\ref{sec:SmEst}, a new approach to this problem, using tools in convex optimization --- we perform a smoothing operation on the convex LSE that retains convexity. Our approach yields an estimator that is (a) smooth, (b) convex and (c) is uniformly close (up to any desired precision) to the LSE. In fact, we can provide theoretical bounds on the quality of the smooth approximation. We emphasize that our approach is very different from the usual techniques (e.g., kernel smoothing) employed in nonparametric statistics to obtain smooth approximations of non-smooth estimators. 	In fact, this technique of smoothing is quite general and applies well beyond the specific instance of piecewise affine functions arising in the context of convex LSE.  
	
{\bf {Lipschitz Regularization:}} In the presence of a limited sample size the convex LSE $\hat \phi_n$ may lead to overfitting, especially towards the boundary of the convex hull of the covariate domain.  This is due to the fact that the fitted subgradients can take very large values near the boundary. To ameliorate this problem we propose least squares estimation of the convex regression function with a (user specified) bounded Lipschitz norm and derive the statistical rates of convergence of this estimator in Section~\ref{sec:unif-Lip-bound1}. The Lipschitz convex LSE thus obtained has lower prediction error and risk when compared to the usual convex LSE. We discuss data-dependent ways to estimate the tuning parameter (i.e., the Lipschitz norm) in practice.
	
{\bf{Additional Structure:}} In several applications of interest, e.g., while estimating production functions and supply/demand functions, it is natural to impose the requirement that the function not only be convex/concave, but that it also be non-decreasing/non-increasing. Our algorithmic framework can be easily adapted to consider such variants of convex regression. We study such variants of our algorithm in Section~\ref{multi-cwise-mono-1}.

{\bf{Experiments:}} We perform a detailed simulation study in Section~\ref{sec:NumExp} demonstrating the superior scalability and performance of our framework over existing publicly available algorithms for this problem. We also consider a few real data examples and highlight the usefulness of convex regression.  
\paragraph{Broader Outlook and Context:} Our paper is intended to be  viewed as a contribution to the larger literature on function estimation in the presence of ``shape constraints''. The earliest (and most extensively studied) such problem is that of isotonic regression (in which the regression function  is presumed to be monotone); see e.g.,~\cite{Brunk55}, and~\cite{AyerEtAl55}. Recently there has been an upsurge of interest in these shape constrained problems, especially in the multivariate setting; see e.g., \cite{CuleEtAl10} and \cite{SW10} for developments in the context of density estimation, and \cite{SS11} and \cite{HD13} for developments in the regression context.


\section{Algorithm for Multivariate Convex Regression}\label{sec:AlgoCvxReg}
In this section we investigate a scalable algorithm for computing the LSE defined in~\eqref{eq:CvxLSE}. We first give a brief discussion as to why Problems~\eqref{eq:CvxLSE} and~\eqref{eq:LSE} are equivalent. Observe that any solution $\B\xi^*_i$'s and $\B\theta^*$ of Problem~\eqref{eq:CvxLSE} can be extended to a convex function by the interpolation rule~\eqref{eq:MaxLSE}. Note that $\hat{\phi}_n(\M{x})$ thus defined is convex in $\Re^d$ and has the same loss function as the optimal objective value of Problem~\eqref{eq:CvxLSE}.
Thus solving Problem~\eqref{eq:CvxLSE} is equivalent to solving~\eqref{eq:LSE}.

We will employ a prototypical version of the alternating direction method of multipliers (ADMM) algorithm; see~\cite{bertsekas-99,Boyd}. For this we consider the following equivalent representation for~\eqref{eq:CvxLSE}:
\begin{equation}\label{eq:CvxLSE-admm1}
\begin{aligned}
\mini_{\B\xi_1, \ldots, \B\xi_n;  \B\theta; \B{\eta}} & \;\;  \;\; \frac{1}{2}\| \M{Y} -  \B{\theta} \|_2^2\\
s.t. \;\;\; &\;\;\;  \eta_{ij} =  \theta_{j} + \langle \B{\Delta}_{ij}, \B{\xi}_{j} \rangle -  \theta_{i}; \qquad \eta_{ij} \leq 0 ; \qquad i , j = 1, \ldots, n,
\end{aligned}
\end{equation}
where $\B\eta=((\eta_{ij})) \in \Re^{n \times n}$ is a matrix with $(i,j)$'th entry $\eta_{ij}$. Define the augmented Lagrangian corresponding to the above formulation as
\begin{eqnarray} 
{\mathcal L}_\rho((\B\xi_1, \ldots, \B\xi_n;  \B\theta; \B{\eta}); \B\nu) & := & \frac{1}{2}\| \M{Y} -  \B{\theta} \|_2^2 + \sum_{i,j} \nu_{ij}  \left ( \eta_{ij}  -  (\theta_{j} + \langle \B{\Delta}_{ij}, \B{\xi}_{j} \rangle -  \theta_{i}) \right) \nonumber\\
&  & \;\;\;\;\;\;+ \; \frac{\rho}{2}  \sum_{i,j} \left( \eta_{ij}  -  (\theta_{j} + \langle \B{\Delta}_{ij}, \B{\xi}_{j} \rangle -  \theta_{i})\right)^2 \label{Lag}
\end{eqnarray}
where $\B\nu \in \Re^{ n \times n}$ is the matrix  of dual variables. We will employ a multiple-block version of 
ADMM for the above problem following~\cite{Boyd}. This requires the following sequential updates as described in Algorithm~\ref{algo1-admm}.
\begin{algorithm}[h!]
Initialize variables $(\B\xi^{(1)}_1, \ldots, \B\xi^{(1)}_n)$,  $\B\theta^{(1)}$, $\B{\eta}^{(1)}$ and  $\B\nu^{(1)}$. 

Perform the following Steps~1---4 for $k \geq 1$ till convergence.
\begin{enumerate}
\item[\bf 1.] Update the subgradients $(\B\xi_1, \ldots, \B\xi_n)$:
\begin{equation}\label{admm-updxi1}
(\B\xi^{(k+1)}_1, \ldots, \B\xi^{(k+1)}_n) \in \argmin_{\B\xi_1, \ldots, \B\xi_n} {\mathcal L}_\rho \left((\B\xi_1, \ldots, \B\xi_n;   \B\theta^{(k)}; \B{\eta}^{(k)}); \B\nu^{(k)} \right).
\end{equation} 
\item[\bf 2.] Update the function values $\B\theta$:
\begin{equation}\label{admm-updtheta1}
\B\theta^{(k+1)} \in \argmin_{\B\theta} {\mathcal L}_\rho \left((\B\xi^{(k+1)}_1, \ldots, \B\xi^{(k+1)}_n;   \B\theta ; \B{\eta}^{(k)}); \B\nu^{(k)} \right).
\end{equation} 
\item[\bf 3.] Update the residual matrix $\B\eta$:
\begin{equation}\label{admm-updeta1}
\B\eta^{(k+1)} \in \argmin_{\B\eta\;:\; \eta_{ij} \leq 0 , \; \forall i, j } {\mathcal L}_\rho \left((\B\xi^{(k+1)}_1, \ldots, \B\xi^{(k+1)}_n;   \B\theta^{(k+1)} ; \B{\eta}); \B\nu^{(k)} \right).
\end{equation} 
\item[\bf 4.] Update the dual variable:
{\small  \begin{equation}\label{admm-updnu1}
\nu_{ij}^{(k+1)} \leftarrow \nu^{(k)}_{ij} + \rho \left ( \eta^{(k+1)}_{ij}  -  \left (\theta^{(k+1)}_{j} + \langle \B{\Delta}_{ij}, \B{\xi}^{(k+1)}_{j} \rangle -  \theta^{(k+1)}_{i}\right ) \right );i, j = 1, \ldots, n.
\end{equation}}
\end{enumerate}
\caption{Multiple-block splitting ADMM for~\eqref{eq:CvxLSE-admm1}}\label{algo1-admm}
\end{algorithm}

\subsection{ADMM with three blocks: Algorithm~1}
We now describe the computations of the different steps appearing in Algorithm~\ref{algo1-admm}.

\subsubsection{Performing the Updates of Algorithm~1}

\subsubsection*{Updating the subgradients: solving Problem~\eqref{admm-updxi1}}
Problem~\eqref{admm-updxi1} is an unconstrained QP in $nd$ variables, where the variables are $\B\xi_{j} \in \Re^d, j = 1, \ldots, n$. Solving such a problem naively has complexity $O(n^3d^3)$. However,  the problem is separable in $\B\xi_{j}$'s --- hence it suffices to consider the updates for each $\B\xi_{j}$  asynchronously, 
over different $j$'s. For every $j$, the subgradient vector $\B\xi_{j}$ can be computed as follows:
%
%
\begin{equation}\label{admm-updxi1-simp}
\hat{\B\xi}_{j} \in \argmin_{\B\xi_{j}} \sum_{i=1}^{n}   \left ( \bar{\eta}_{ij} - \langle \B{\Delta}_{ij}, \B{\xi}_{j} \rangle  \right)^2\; {\small \imply\;  \hat{\B\xi}_{j} = \Big(\sum_{i} \B\Delta_{ij}\B\Delta_{ij}^\top\Big)^{-1} \Big(  \sum_{i} \B\Delta_{ij} \bar{\eta}_{ij} \Big)}
\end{equation}
where $\bar{\eta}_{ij} =  {\nu_{ij}}/{\rho} + \eta_{ij} - (\theta_{j} - \theta_{i})$, which is a least squares regression problem for every $j=1, \ldots, n$. If $n>d$ and the design points come from a continuous distribution,
$\hat{\B\xi}_{j}$ is unique with probability one. We note that the matrices $\overline{\Delta}_{j}:= \left(\sum_{i} \B\Delta_{ij}\B\Delta_{ij}^\top\right)^{-1} , j = 1,\ldots, n$, need not be computed at every iteration, they can be computed offline at the onset of the algorithm:
computing $\sum_{i} \B\Delta_{ij}\B\Delta_{ij}^\top$ takes $O(nd^2)$ and the inversion $O(d^3)$ for every $j$.  Thus, the total cost for all $j$ values would be $O(n^2d^2 + nd^3)$, a computation that can be done once in parallel at the onset of the algorithm. Once the inverses are computed, computing $\hat{\B\xi}_{j}$ for all $j$ requires an additional cost of $O(n^2d + nd^2)$ --- this is due to the cost of computing the $\bar{\eta}_{ij}$'s,  the matrices  $\sum_{i} \B\Delta_{ij} \bar{\eta}_{ij}$ and the subsequent matrix multiplications leading to $\hat{\B\xi}_{j}$.  Since we typically have $d \ll n$, the cost per iteration is $O(n^2)$.


\subsubsection*{Updating the function values: solving Problem~\eqref{admm-updtheta1}}
We introduce the following notations:
\begin{equation*}
\begin{aligned}
\{\M{D}\B\theta\}_{(i - 1)n + j }  = (\theta_{j} - \theta_{i}),\;\;
\{\mathrm{vec}({\B\nu})\}_{(i - 1)n + j }  =  \nu_{ij}, \;\;
\{\mathrm{vec}({\widetilde{\B\eta}})\}_{(i - 1)n + j }  =  (\eta_{ij} - \langle \B{\Delta}_{ij}, \B\xi_{j} \rangle),\\
 \end{aligned}
 \end{equation*}
for $i,j =1, \ldots, n$, where $\M{D}$ is a sparse $n^2 \times n$ matrix.
Using the above notation, the optimization problem~\eqref{admm-updtheta1} reduces to the minimization of the following function (with respect to $\B\theta$): 
\begin{equation*}
\frac{1}{2}\| \M{Y} -  \B{\theta} \|_2^2 - \left\langle \mathrm{vec}{(\B\nu)}, \M{D} \B\theta \right\rangle  + \frac{\rho}{2} \left\| \mathrm{vec}(\tilde{\B\eta}) - \M{D} \B\theta \right\|_{2}^2,
\end{equation*}
which is equivalent to solving the following system:  
\begin{equation}\label{upd-theta-ls1}
 (\M{I} + \rho \M{D}^\top\M{D})\hat{\B\theta}  = \underbrace{\M{Y} + \M{D}^\top \mathrm{vec}{(\B\nu)} + \rho \M{D}^\top \mathrm{vec}{(\tilde{\B\eta})}}_{:=\M{v}}.\vspace{-0.05in}
 \end{equation}

Computing the vector $\M{v}$ will cost $O(n^2)$ flops.
The matrix appearing on the left-hand side of the above equation, $(\M{I} + \rho \M{D}^\top\M{D})$, has dimension $n \times n$. A direct inversion of 
the matrix to solve for $\B\theta$ will have a complexity of $O(n^3)$. We show below that the cost can be reduced substantially by exploiting the structure of the linear system~\eqref{upd-theta-ls1}. From the definition of  $\M D$ we have 
 $\M D^\top \M D = 2n \M I_{n \times n} - 2 \; \M 1 \M1^\top,$
 where $\M 1$ denotes the $n \times 1$ vector of all ones and $\M I_{n \times n}$ is the identity matrix of order $n$. Thus we have
$(\M{I} + \rho \M{D}^\top\M{D}) = (1+2n\rho)\M{I} - 2\rho\M{1}\M{1}^\top$ and 
\begin{align}\label{rep-IDD}
(\M{I} + \rho \M{D}^\top\M{D})^{-1}= \left(\frac{1}{1+2n\rho}\M{I} + \frac{2\rho}{1+2n\rho}\M{1}\M{1}^\top\right),
\end{align}
using which the system~\eqref{upd-theta-ls1} can be solved for $\B\theta$:
\begin{equation}
\begin{aligned}
\;\;\;& \hat{\B\theta}  =&  \left(\frac{1}{1+2n\rho}\M{I} + \frac{2\rho}{1+2n\rho}\M{1}\M{1}^\top\right) \M{v} 
\imply \;\;\;& \hat{\theta}_i =& \frac{1}{(1 + 2n \rho)} \left( v_{i} + 2\rho \sum_{i} v_{i}    \right).
\end{aligned}
\end{equation}
The above computation can be done quite efficiently in $O(n)$ flops, given $\M{v}$. 
Thus the total cost per iteration in computing $\hat{\B\theta}$ is $O(n^2)$.


\subsubsection*{Updating the residuals: solving Problem~\eqref{admm-updeta1}}
Updating the residuals, i.e., $\eta_{ij}$'s, are simple --- due to the separability of the objective function the partial minimization splits into $n^2$ independent univariate optimization problems. It suffices to consider the following update rule for any $(i,j)$:
\begin{equation}\label{upd-nu-closed1}
\hat{\eta}_{ij} = \min \left \{\theta_{j} + \langle \B{\Delta}_{ij}, \B{\xi}_{j} \rangle -  \theta_{i} - \frac{1}{\rho} \nu_{ij} , 0 \right \} ; \;\;\;
i, j = 1, \ldots, n. 
\end{equation}
Note that the inner products $\langle \B{\Delta}_{ij}, \B{\xi}_{j} \rangle$ are already available as a by-product of
the $\B\theta$ update, while solving Problem~\eqref{admm-updtheta1}; hence they need not be computed from scratch. 
Thus, the cost for performing this operation is $O(n^2)$, since there are $n^2$ many variables to be updated.

\paragraph{Computational Complexity:} 
Note that updating the dual variable $\B\nu$ using~\eqref{admm-updnu1} requires $O(n^2)$ flops. Thus, gathering the discussion above, the cost per iteration of Algorithm~1 is $O( \max \{ n^2d, nd^3 \} )$, with an additional $O(n^2d^2+nd^3)$ for the offline 
computation of matrix inverses for Problem~\eqref{admm-updxi1}.
If $d \ll n$, the total cost is $O(n^2)$, which seems reasonable since the optimization problem involves $O(n^2)$ constraints.

\subsubsection{Optimality Conditions for Problem~\eqref{eq:CvxLSE}}
Let $\B\theta^*, \eta_{ij}^*$ and $\B\xi^*_{j}$, for $i, j \in \{ 1, \ldots, n\}$, denote optimal solutions to Problem~\eqref{eq:CvxLSE}, or equivalently, Problem~\eqref{Lag}, and 
let $\B\nu^*$ denote an optimal dual variable.
The optimality conditions for the problem~\citep{BV2004,Boyd} are given by:
\begin{subequations}\label{opt-conds-1}
\begin{align}
\eta^*_{ij}  - \left( \theta_{i}^* + \langle \B \Delta_{ij}, \B\xi_{j}^*\rangle - \theta_{j}^* \right) =&\; 0,~~~~~~~i,j = 1, \ldots, n, \label{cond-11} \\
\sum_{i=1}^{n} \nu_{ij}^*\Delta_{ij}  =&\; 0,~~~~~~~j = 1, \ldots, n, \label{cond-12}\\
(  \B\theta^* - \M{Y}) - \M{D}^\top \text{vec}(\B\nu^*) =&\; \M{0},~~~~~~~\label{cond-13}\\
\eta^*_{ij} - \min \left \{ \eta_{ij}^* - \frac{1}{\rho} \nu^*_{ij} , 0 \right \} =&\; 0,~~~~~~~~i,j = 1, \ldots, n\label{cond-14},
\end{align}
\end{subequations}
where~\eqref{cond-11} is the primal feasibility condition,~\eqref{cond-12} is the optimality condition with respect to~$\B\xi_{j}$'s,~\eqref{cond-13} is the optimality condition with respect to~$\B\theta$ and~\eqref{cond-14} denotes the optimality condition with respect to $\B\eta$.

\subsection{Convergence properties}
Algorithm~\ref{algo1-admm} is a direct application of the schematic multiple-block (with three blocks) ADMM algorithm described in~\cite{Boyd} --- it is quite simple and easy to implement. There is, however, one caveat. As soon as the number of blocks becomes larger than two, the multiple block version of ADMM however, does not necessarily converge --- see e.g., ~\cite{chen2014direct} for a counter example. Multiple block versions of ADMM, however, can be shown to be convergent, under certain restrictive assumptions on the problem and by possibly modifying the algorithm; see e.g.,~\cite{hong2012linear}. 
While a thorough convergence analysis of multiple block ADMM is not the main theme or focus of this paper, it is indeed quite simple to check if Algorithm~1 has converged or not by verifying if the conditions of optimality~\eqref{opt-conds-1} are met. In all of our experiments, Algorithm~1 was  indeed found to converge.
Nevertheless, we discuss a simple variant of Algorithm~1 that has superior convergence guarantees. 
The main principle behind the modification(s) is quite simple: since a three block version of the ADMM is not guaranteed to converge, 
one can resort to a two-block version of the ADMM. One way to obtain a two-block version of ADMM is by combining the serial updating of $\B\theta$ and $\B\eta$, i.e., Problems~\eqref{admm-updtheta1} and~\eqref{admm-updeta1} into one that 
jointly optimizes over both $\B\theta$ and $\B\eta$, i.e., 
\begin{equation}\label{upd-algo2}
\begin{aligned}
(\B\theta^{(k+1)}, \B\eta^{(k+1)}) \in & \argmin_{\B\theta, \B\eta:\; \eta_{ij} \leq 0, \forall  i, j}& {\mathcal L}_\rho \left((\B\xi^{(k+1)}_1, \ldots, \B\xi^{(k+1)}_n;   \B\theta ; \B{\eta}); \B\nu^{(k)} \right).
\end{aligned}
\end{equation}
The update in $\B\xi_{j}$'s and the dual variable update remains the same. Convergence guarantees of the resultant two block variant of Algorithm~1 are described in~\cite{Boyd}.
Note that update~\eqref{upd-algo2} can be performed by using block coordinate descent (see e.g.,~\cite{bertsekas-99} and \cite{FHT2007})  on the blocks $\B\theta$ and $\B\eta$, 
which is essentially equivalent to performing updates~\eqref{admm-updtheta1} and~\eqref{admm-updeta1} sequentially till convergence. 

%
%
%

If we collapse all the variables $(\B\xi_{1}, \ldots, \B\xi_{n}, \B\theta, \B\eta)$ into one block  $\B\zeta$ (say) then Problem~\eqref{Lag} may be thought of as an  augmented Lagrangian in the variable $\B\zeta$, for which one may apply the augmented Lagrangian Method of Multipliers --- a method for which convergence is relatively well understood (\cite{bertsekas-99}); also see the recent work of~\cite{aybat2012first}.
We briefly describe this method below.


\noindent {{\bf Algorithm~2}}
\begin{enumerate}
\item Update:  $\B\zeta^{(k+1)}:=\left(\B\xi^{(k+1)}_{1}, \ldots, \B\xi^{(k+1)}_{n}, \B\theta^{(k+1)}, \B\eta^{(k+1)} \right)$ as:
\begin{equation}\label{upd-joint-alg3}
\B\zeta^{(k+1)} \in \argmin_{\B\xi_{1}, \ldots, \B\xi_{n}, \B\theta, \B\eta} {\mathcal L}_{\rho}(\B\zeta; \B\nu^{(k)}) \;\; s.t. \;\; \eta_{ij} \leq 0, \;\forall \; i,j=1, \ldots, n.
\end{equation}
\item Update: $\B\nu^{k+1}$ as in~\eqref{admm-updnu1}.
\end{enumerate}


We make some remarks about Algorithm~2 below.


$\bullet$ The \emph{joint} optimization~\eqref{upd-joint-alg3} with respect to $\B\zeta$ can be done via block coordinate descent by updating $\B\xi_{i}$'s, $\B\theta$ and $\B\eta$ sequentially as per 
the update rules~\eqref{admm-updxi1},~\eqref{admm-updtheta1} and~\eqref{admm-updeta1}.

$\bullet$ The bottleneck in Algorithm~2 is the optimization Problem~\eqref{upd-joint-alg3}, since it needs to be performed for every iteration $k$.  The problems, however, can be warm-started by using the estimates from the previous iteration for the current iteration.
 Since Algorithm~1 has excellent practical performance, i.e., the block updates are fairly cheap to carry out, we found it useful to initialize Algorithm~2 with a solution obtained from Algorithm~1. This hybrid version enjoys the convergence guarantees of Algorithm~2 as well as the good practical performance of Algorithm~1;  see Section~\ref{sec:NumExp} where our experimental results are described.

$\bullet$ The step-size $\rho$ can be updated \emph{dynamically} and the optimization subproblems~\eqref{upd-joint-alg3} can be solved \emph{inexactly} such that the overall algorithm converges to an $\epsilon$-optimal and $\epsilon$-feasible solution in  $O(\frac{1}{\epsilon})$ operations, as shown in~\cite{aybat2012first}.

\section{Smoothing Non-smooth Convex estimators}\label{sec:SmEst}
For simplicity of notation, in this section, we will write $\hat{\phi}_n$ as $\hat{\phi}$. Problem~\eqref{eq:CvxLSE} describes a method to estimate the unknown convex regression function at the given covariate values $\M{X}_{i}$'s. Although~\eqref{eq:MaxLSE} describes a way of extending it to the whole of $\Re^d$, the obtained estimator is neither smooth nor unique (as $\hat{\B \xi_j}$'s are not unique). While there are several ways in which the LSE can be defined beyond the data points, we briefly describe below a natural alternative that produces a unique estimator.   

Let $\X := \{\M{X}_1,\ldots, \M{X}_n\}$ and let $\text{Conv}(A)$ denote the convex hull of a set $A$. We can take the nonparametric LSE to be the function $\check{\phi}:\Re^d\rightarrow \Re$ defined by
\begin{equation}\label{eq:CanonLSE}
\check{\phi}(\M{x}) := \inf\left\{ \sum_{k=1}^n \alpha_k \hat \theta_k : \sum_{k=1}^n \alpha_k = 1,\ \sum_{k=1}^n \alpha_k \M{X}_k = \M{x},\ \alpha_k \geq 0 \ \forall \; k= 1,\ldots, n\right\}
\end{equation}
for any $\M{x} \in \Re^d$. Here we take the convention that $\inf(\emptyset) = +\infty$. The function $\check{\phi}$ is well-defined and is finite on $\text{Conv}(\X)$. In fact, $\check{\phi}$ is a polyhedral convex function (i.e., a convex function whose epigraph is a polyhedron; see \cite{SS11}). It is easy to see that $\check{\phi}$ is unique (as it only depends on the $\theta_k$'s); in fact, $\check{\phi}$ is the largest convex function satisfying the constraints $\check{\phi}(\M{X}_k ) = \theta_k$, for $k=1,\ldots, n$; see \cite{SS11}.

We describe below a novel approach to obtaining smooth approximations to the convex regression estimators $\hat{\phi}$ and $\check{\phi}$ with provable theoretical bounds on the quality of approximation. Our approach is different from the usual methods in nonparametric statistics to obtain smooth estimators under known shape restrictions; see e.g.,~\cite{M88},~\cite{M91},~\cite{BD07},~\cite{AguileraEtAl11},~\cite{Du13} and the references~therein. 




We propose a method of finding a smooth approximation $\phi^{\text{Sm}}$ to the piecewise affine LSE $\hat{\phi}$ (or $\check{\phi}$)
with the following properties:
\begin{enumerate}
\item $\phi^{\text{Sm}}$ is differentiable and its gradient $ \nabla \phi^{\text{Sm}}(\cdot)$ is Lipschitz continuous with parameter $\ML$, i.e., 
$ \| \nabla \phi^{\text{Sm}}( \M{x}_{1} ) - \nabla \phi^{\text{Sm}}(\M{x}_{2}) \|_2  \leq \ML \| \M{x}_{1} -\M{x}_{2} \|_2 $
for $\M{x}_{1}, \M{x}_{2} \in \Re^{d}$.
\item $\phi^{\text{Sm}}$ is uniformly close to $\hat{\phi}$ in the following sense:
$ \sup_{\M{x}} | \phi^{\text{Sm}}(\M{x}) - \hat{\phi} (\M{x}) | \leq \tau (\ML),$
where, $\tau (\ML)$ is a function of the global Lipschitz gradient parameter $\ML$.
\end{enumerate}
In the next section we briefly describe the general technique for smoothing non-smooth functions and then in Section~\ref{sec:SmLSE} (and Section~\ref{sec:SmLSE2}) we specialize to the case of $\hat{\phi}$ (and $\check{\phi}$).

\subsection{Smoothing structured convex non-smooth functions}\label{sec:smoothing1}

\paragraph{Preliminaries.} We start with some notation. Consider $\Re^d$ with a norm $\|\cdot \|$ and denote its dual norm by
$ \| \M s \|^* := \; \max_{\M z:  \|\M z\| = 1} \langle \M s, \M z \rangle,$ where $\langle \cdot, \cdot \rangle$ denotes the usual inner product.
\begin{mydef}\label{defn-1}
We say that a function $f : \Re^d \to \Re$ is \emph{smooth} with parameter $\ML$ if it is continuously differentiable and  its gradient is Lipschitz with parameter $\ML$, i.e.,
$\| \nabla f(\M x)  - \nabla f( \M y ) \| \leq \ML \| \M x- \M y \|$ for all $\M x, \M y \in \Re^d$.
\end{mydef}

The framework described below follows the methodology introduced by~\cite{Nesterov05} who describes an elegant smoothing procedure for the minimization of non-smooth functions having favorable geometrical structures.
For the sake of completeness we present the general framework below.  
We consider the spaces $\Re^q$ and $\Re^p$ with norms $\| \cdot\|_{\#}$ and $ \|\cdot \|_{\dagger} $ respectively. 
For a matrix $A_{p \times q}$, define its matrix norm $\| A\|_{\#, \dagger }$  induced by the norms 
$\| \cdot\|_{\#}, \|\cdot \|_{\dagger} $ as 
$$ \| A\|_{\#, \dagger } := \max_{\M u ,\M v }  \left \{ \langle A \M u , \M v  \rangle  :   \| \M u \|_{\#}  = 1,  \|\M  v\|_{\dagger} = 1, \;\M u \in \Re^{q} ,  \M v \in \Re^{p} \right\} $$ 
where $\langle \cdot, \cdot \rangle$ denotes the usual Euclidean inner product. It is easy to see that 
$$ \| A\|_{\#, \dagger } = \| A^\top \|_{\dagger, \# } = \max_{\M u} \{ \| A \M u \|^*_{\dagger} : \| \M u \|_{\#} = 1 \} =  
\max_{\M v} \{ \| A^\top \M v \|^*_{\#} : \| \M v \|_{\dagger} = 1 \}.$$ Let  $Q \subset \Re^{p}$ be a closed convex set and $\rho(\cdot)$ be a proximity (prox) function (see e.g., \cite{Nesterov05,nesterov2004introductorynew}) of the set $Q$. We will assume that $\rho(\cdot)$ is continuously differentiable and strongly convex on $Q$ (with respect to the norm $\|\cdot \|_{\dagger}$) with strong convexity parameter one, i.e., 
$$ \rho(\M u_{1}) \geq  \rho(\M u_{2}) + \langle \nabla \rho(\M u_{2}) ,\M u_{2} - \M u_{1} \rangle + \frac12\| \M u_{2} - \M u_{1}  \|_{\dagger}^2$$
for any $\M u_{1}, \M u_{2} \in Q$. In particular, this implies that if $\M w_{0} \in \argmin_{\M w \in Q} \rho(\M w)$ then for any $\M w \in Q$, we have $$ \rho(\M w) - \rho(\M w_{0})  \geq \frac{1}{2}\| \M w - \M w_{0} \|_{\dagger}^2. $$ Without loss of generality we can take the prox(imity) center $\M w_{0}$ to satisfy $\rho(\M w_{0})=0$.

Consider a function $\gamma: \Re^q \times (0,\infty) \to \Re$   given by \begin{equation}\label{gen-smooth-max-1}
\gamma(\M z ; \tau) := \max_{\M w \in Q} \;\left \{ \langle A \M z, \M w  \rangle - \tau \rho(\M w) \right \}
\end{equation}
where $\tau > 0 $ is a regularization parameter.  To motivate the reader, let us consider a simple example with $Q$ given by the unit $\ell_{1}$-ball, i.e., $Q= \{\M  w : \| \M w \|_{1} \leq 1 \}$. Note that if $\tau =0$ and $Q$ as described above, $\gamma(\M z ; 0)= \| A \M z \|_{\infty}$ which is a non-smooth function.  The non-smoothness arises precisely due to the fact that $\hat{\M w} \in \argmax_{\M w \in Q}   \langle A \M z, \M w  \rangle $ is not unique.  
One can get rid of the non-smoothness by using the modification suggested in~\eqref{gen-smooth-max-1}. Since the optimization problem in~\eqref{gen-smooth-max-1} involves the maximization of a strongly  concave function over a closed convex set, its maximum is attained and is unique. Thus the presence of the perturbation term $\tau \rho(w)$ with $\tau >0$ has interesting consequences --- it makes the function $\M z \mapsto \gamma(\M z ; \tau)$ smooth (as in Definition~\ref{defn-1}) and the amount  of smoothness imparted via the regularization  can be precisely quantified. In this vein, we have the following theorem.
\begin{lem}\label{main-thm-nest-1}
For any fixed $\tau >0$, the function $\gamma(\M z; \tau)$, defined in~\eqref{gen-smooth-max-1}, is differentiable in $\M z$ and its 
gradient is given by\footnote{$\nabla_{1} \gamma(\M z; \tau)$ refers to the partial derivative of $\gamma(\M z; \tau)$ with respect to $\M z$.}
$\nabla_{1} \gamma(\M z; \tau)  =   A^\top\hat{\M w}^{\tau},$ 
where  $\hat{\M w}^\tau \in \argmax_{\M w \in Q}  \left\{\langle A \M z, \M w  \rangle - \tau \rho(\M w)\right\}$. Furthermore, the gradient map $\M z \mapsto \nabla_{1} \gamma(\M z; \tau)$ is Lipschitz continuous  with
parameter $\frac{\|A\|_{\#, \dagger}^2}{\tau}$.
\end{lem}

We also have the following bounds describing how close $\gamma(\M z; \tau)$ is to the unperturbed function  $\gamma(\M  z; 0)$: 
\begin{equation}\label{lbdiff-1}
\begin{aligned}
 \gamma(\M z ; \tau)
& \geq &  \sup\limits_{\M w \in Q}  \langle A\M z, \M w \rangle - \tau \sup\limits_{\M w \in Q} \rho(\M w) 
& = &  \gamma(\M z; 0)  - \tau \sup\limits_{\M w \in Q} \rho(\M w),
 \end{aligned}
 \end{equation}
 and
\begin{equation}\label{ubdiff-1}
 \gamma(\M z ; \tau) = \sup_{\M  w \in Q }  \{ \langle A\M z, \M w \rangle - \tau \rho(\M w)\}  \leq \sup_{\M w \in Q}   \langle A\M z, \M w \rangle =  \gamma(\M z ; 0),
 \end{equation}
where the last inequality follows as a consequence of the non-negativity of the prox function. 
\begin{lem}\label{lem-unif-bound1}
For any $\tau \geq 0$,  the perturbation $ \gamma(\M z ; \tau)$  of $ \gamma(\M z ; 0)$ satisfies the following uniform bound over $\M z$:
\begin{equation}\label{eq:Bound}
 \gamma(\M z ; 0)   - \tau \sup_{\M w \in Q} \rho(\M w) \leq  \gamma(\M z ; \tau)  \leq \gamma(\M z ; 0) .
\end{equation}
\end{lem}
We present below a summary of the smoothing procedure described above:
\begin{itemize}
\item The function $\M z \mapsto  \gamma(\M z ; 0)$ may be non-smooth in $\M z$. For $\tau >0$, $\M z  \mapsto\gamma(\M z ; \tau)$ is smooth~(by Theorem~\ref{main-thm-nest-1}), convex and has Lipschitz continuous gradient with parameter $L=O(1/\tau)$. 
 
\item The smooth function $\gamma(\M z; \tau)$ serves as a uniform approximation to $\gamma(\M z; 0)$; the quality of approximation is quantified in Lemma~\ref{lem-unif-bound1}.  The approximation error is (upper) bounded by $ \tau \sup_{\M w \in Q} \rho(\M w) $ --- a quantity that depends on $Q$ and the choice of the function $\rho(\cdot)$.  

\item For a given $\rho(\cdot)$ and $Q$, the smoothness of the function $\M z \mapsto \gamma(\M z; \tau)$ is  $O(1/\tau)$, which is inversely related to the 
order  of approximation error, given by, $O(\tau)$.
\end{itemize}
\subsection{Smooth post-processing of the convex LSE}\label{sec:SmLSE}
We will use the above framework to smooth the convex function estimators. Consider first, the piecewise affine estimator $\hat{\phi}$. For $\M x \in \Re^d$, we can always represent $\hat{\phi}$ as defined in~\eqref{eq:MaxLSE} as:
\begin{equation}\label{pwise-affine-gen1}
 \hat{\phi}(\M{x}) = \max \left\{ \M a^\top_{1}\M x + b_{1} , \ldots, \M a^\top_{m}\M x + b_{m} \right\}.
\end{equation}
The special case of $\hat{\phi}$ as defined in~\eqref{eq:MaxLSE} can be expressed with
$\M{a}_{i},b_{i}$'s, for $i=1, \ldots, m,$ with $m = n$, chosen as:
\begin{equation}\label{aibi-choice}
\M{a}_{i} = \hat{\B\xi}_{i}, \;\;\;\;b_{i} = \hat{\theta}_{i} - \langle \hat{\B\xi}_{i}, \M{X}_{i} \rangle, \;\;\; i = 1, \ldots, m.
\end{equation}

Observe that $\hat{\phi}$ admits the following pointwise representation:
\begin{equation}\label{argmin-fn-1}
\begin{aligned}
\hat{\phi}(\M x) =&\max \limits_{\M w} & \sum\limits_{i=1}^m \;\; w_{i} \left (\M  a_{i}^\top\M x + b_{i}   \right ) & \;\;
&s.t. \;\;& \sum\limits_{i=1}^{m} w_{i} = 1, w_{i} \geq 0, i = 1, \ldots, m.
\end{aligned}
\end{equation}
Let us denote the unit simplex appearing in the constraint set of the above optimization problem by 
$\Delta_{m}:= \left\{ \M w : \sum_{i} w_{i} = 1, \;w_{i} \geq 0, i = 1, \ldots, m\right\}.$ Observe that the non-smoothness in the function $\hat\phi(\cdot)$ arises due to the non-uniqueness of $\hat{\M w}$ where $\hat{\M w}$ is a maximizer of the linear optimization problem appearing in~\eqref{argmin-fn-1}. 

Following the framework developed in the previous section, it is easy to see that $\gamma(\M x; 0)$, as defined in~\eqref{gen-smooth-max-1}, is indeed the function $\hat{\phi}(\M x)$ and $Q$ is $\Delta_{m}$ (with a suitable renaming of variables and adjusted dimensions). The smooth approximation $\tilde \phi$ of the non-smooth function $\hat{\phi}$ is thus obtained by the construction~\eqref{gen-smooth-max-1}:
\begin{equation}\label{add-conv-1}
\begin{array}{l c c l}
\tilde{\phi}(\M  x ; \tau) =& \max \limits_{\M w} & \sum\limits_{i=1}^m \;  w_{i} \left (\M  a_{i}^\top \M x + b_{i}   \right )  - \tau \rho(\M w) &\\
&s.t. & \sum\limits_{i=1}^{m} w_{i} = 1, w_{i} \geq 0, i = 1, \ldots, m,
\end{array}
\end{equation}
where $\rho(\M w)$ is a prox function on the $m$ dimensional unit simplex. In the following we examine the consequences of using two different choices for the prox function.

\subsubsection{Smoothing via the squared error prox function}\label{sm-sq-err-prox1}
A natural choice for smoothing comes from the squared error prox function, i.e., $\rho(\M w) = \frac{1}{2} \left\|\M  w - \frac{1}{m} \M{1} \right\|_{2}^2$ for which both norms $\|\cdot \|_\dagger, \| \cdot \|_{\#}$ (as appearing in Section~\ref{sec:smoothing1}) are taken to be the standard Euclidean norms. The smoothed approximation $\tilde{\phi}(\M x; \tau)$ can be obtained by solving problem~\eqref{add-conv-1} with the squared error prox function. 
The optimization problem in~\eqref{add-conv-1} is equivalent  to the following convex program:
$$\mini_{\M w} \; \; \left\{ \half \sum_{i=1}^m w_{i}^2  -  \sum_{i=1}^m w_{i} \tilde{c}_{i}\right\} \;\;\; s.t. \;\;  \sum_{i=1}^m w_{i} = 1, w_{i} \geq 0 , i = 1, \ldots, m,$$
where $\tilde{c}_{i} = (a_{i}^\top \M x + b_i)/\tau - 1/m $, for all $i = 1, \ldots, m $.
The above problem is exactly equivalent to the Euclidean projection of 
$\tilde{\M c}:= (\tilde{c}_{1}, \ldots, \tilde{c}_{m})$ onto the unit $m$ dimensional simplex. Though there does not exist a closed form expression for this projection operation, it can be computed quite efficiently with complexity $O(m\log m)$ requiring a sorting operation on $m$ numbers; see \cite{michelot1986finite}.
  The approximation error associated with this smooth approximation $\tilde{\phi}(\M x; \tau)$ of $\hat{\phi}(\M x)$ is given by
(see Lemma~\ref{lem-unif-bound1})  
\begin{equation}\label{eq:UnifBd1}
\sup_{\M x \in \Re^d} |\tilde{\phi}(\M x; \tau) -  \hat{\phi}(\M x)| \le  \tau \sup_{\M w \in \Delta_{m}}  \frac{1}{2} \| \M w - \M{1}/m \|_{2}^2 = \tau ( 1 - 1/m)
\end{equation}
and the Lipschitz constant of the gradient is $\|A\|_{2,2}^2/\tau = \lambda_{\max}( A^\top A)/\tau,$ since $\|A\|^2_{2,2}=\lambda_{\max}( A^\top A)$, the maximum eigenvalue of $A^\top A$, where $A$ is the $m \times (d+1)$ matrix whose $i$'th row is $(b_i, \M  a_{i}^\top)$.
 Suppose that one seeks to obtain a smooth approximation $\tilde{\phi}(\M x; \tau)$ to $\hat{\phi}(\M x)$, with approximation error given by $\epsilon$. This leads to a choice of $\tau$ that results in a bound of the Lipschitz constant of the gradient by $\lambda_{\max}( A^\top A) (m - 1)/(m\epsilon).$

\begin{figure}[h!]
\begin{center}
\resizebox{\textwidth}{0.13\textheight}{\begin{tabular}{c cc}
\rotatebox{90}{~~~~~~~~~~~~~~~~~~~~$\hat{\phi}$}&\includegraphics[height=2.2in,width= 2.5in,angle= 0,trim =1.0cm 1.5cm 1.cm 1.8cm, clip = true]{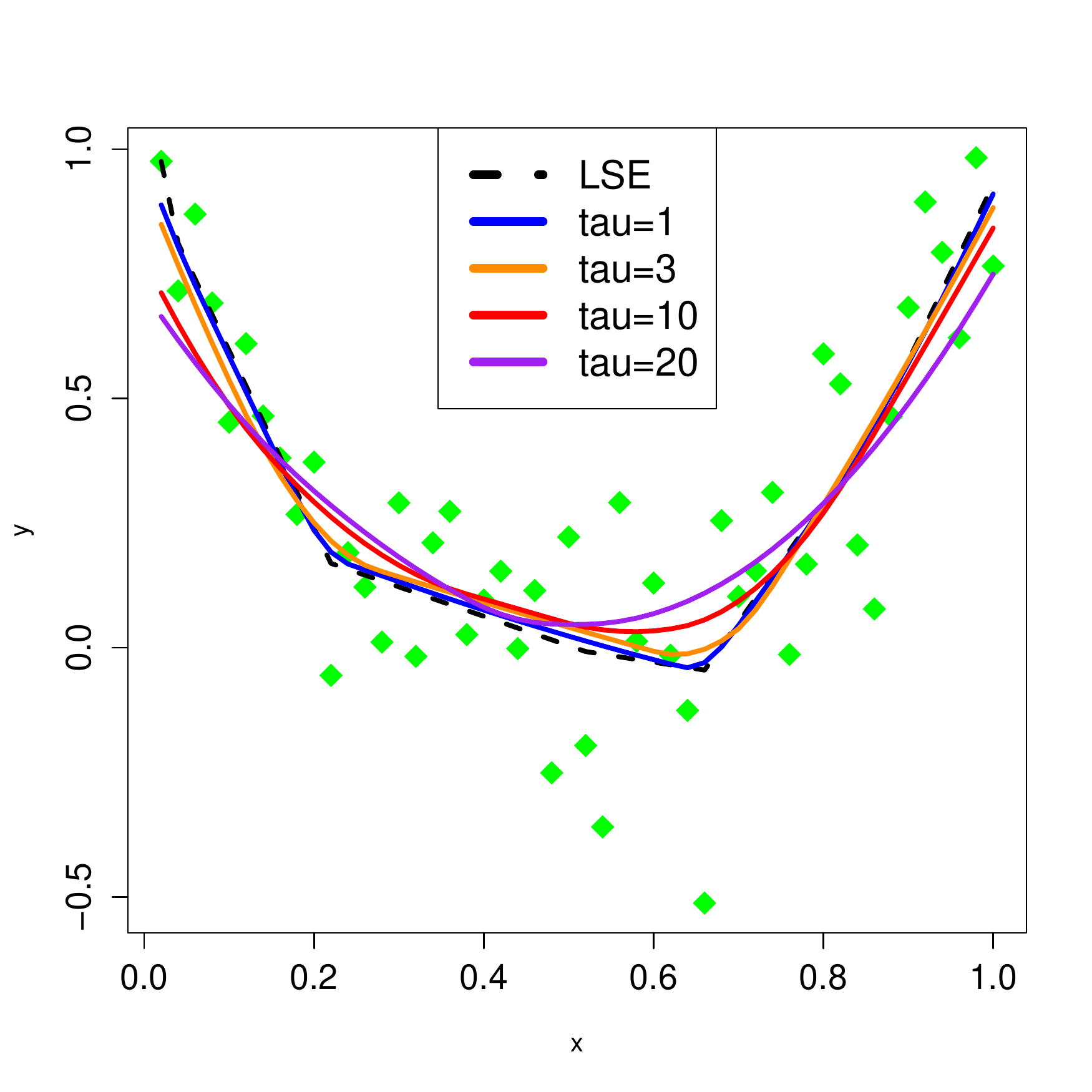} \hspace{3mm} &
\includegraphics[height=2.2in,width= 2.5in,angle= 0,trim =1.0cm 1.5cm 1.cm 1.8cm, clip = true]{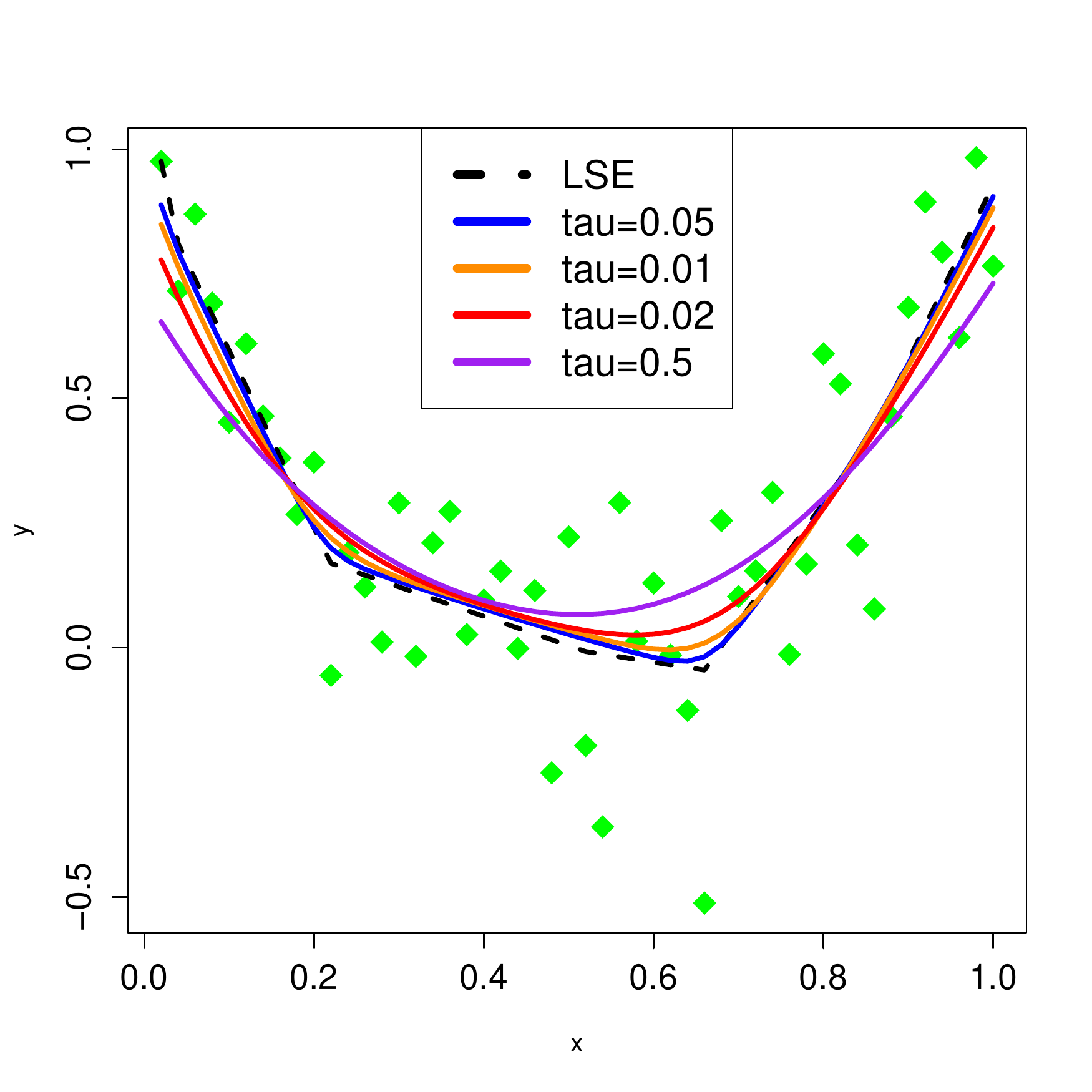}\\
&\sf \scriptsize {$X$} &  \sf \scriptsize{$X$}  \\
\end{tabular}}
\caption{Plots of the data points and the convex LSE $\hat \phi$ with the bias corrected smoothed estimators for four different choices to $\tau$ using the squared error prox function (left panel) and entropy prox function (right panel).}
\label{fig:SmoothPlots}
\end{center}
\end{figure}

\subsubsection{Smoothing via the Entropy prox function}\label{sm-ent-err-prox1}
Let us next consider the entropy prox function on the unit simplex, i.e.,
$\rho(\M w) = \sum_{i=1}^{m} w_{i} \log(w_{i}) + \log m $
and let $\|\cdot \|_{\dagger}$ and $\| \cdot \|_{\#}$ be the 
$\ell_{1}$-norm in $\Re^m$. This prox function is strongly convex with respect to the standard $\ell_{1}$-norm. For a simple proof of this fact, note that for any $\M h \in \Re^{m}$ we have 
$\langle \nabla^2 \rho(\M w) \M h, \M h \rangle   = \sum_{i} h_{i}^2/w_{i}.$  By the Cauchy-Schwarz inequality it follows that 
$ \left (\sum_{i} h_{i}^2/w_{i} \right ) \left( \sum_{i} w_{i} \right)  \geq \left(\sum_{i} |h_{i}| \right)^2 $
which implies strong convexity of the entropy prox function $\rho(\cdot)$ with respect to the $\ell_{1}$-norm. Furthermore, it can be shown that for this choice of prox function
\begin{equation}\label{eq:UnifBd2}
\sup_{\M x \in \Re^d} |\tilde{\phi}(\M x; \tau) -  \hat{\phi}(\M x)|  \le \tau\sup_{\M w \in \Delta_{m}} \rho(\M w) = \tau \log m  
\end{equation}
and the Lipschitz constant of the gradient is given by 
$ \|A\|_{1,1}^2/\tau   = \left( \max_{i,j} | A_{ij} | \right)^2/ \tau.$ Thus, for  an approximation error budget of  $\epsilon$, the corresponding Lipschitz constant of the gradient is given by 
${\left( \max_{i,j} | A_{ij} | \right)^2 \log m}/{\epsilon}.$
For the entropy prox function, there is a simple analytic expression for the smoothed approximation $\tilde \phi(\cdot; \tau)$ of $\hat{\phi}(\cdot)$, given by
\begin{eqnarray*}
\tilde \phi(\M x; \tau) &= & \sup_{\M w \in \Delta_{m}}\left\{\sum_{i=1}^m w_{i} \left(\M a_{i}^\top \M x + b_{i}\right) -  \tau\left(\sum_{i=1}^m w_{i}\log(w_{i}) + \log m  \right) \right\} \\
 &=& \tau \log \left( \sum_{i=1}^m \exp\left(\frac{\M a_{i}^\top \M x + b_{i}}{\tau} \right)  \right)  - \tau \log m.
\end{eqnarray*}

Figure~\ref{fig:SmoothPlots} shows the smoothed convex estimators obtained from the two procedures (after a bias correction, described in Section~\ref{sec:BiasCorr}) for different values of $\tau$ when $n=50$ and $d=1$. We see that both smoothing methods yield similar results and can produce estimators with varying degrees of smoothness. A similar phenomenon is observed when $d$ exceeds 1.

The results described in Section~\ref{sm-sq-err-prox1} and Section~\ref{sm-ent-err-prox1} lead to the following theorem.
 
 \begin{thm}\label{thm-approx-1}
Let $\tilde{\phi}(\M{x} ; \tau)$, as defined via~\eqref{add-conv-1}, be a smoothed approximation of the piecewise affine convex LSE fit $\hat{\phi}(\M x)$ as in~\eqref{pwise-affine-gen1} such that
the following holds:
$$\sup_{\M x \in \Re^d} |\tilde{\phi}(\M x; \tau) -  \hat{\phi}(\M x)| \le \epsilon,$$
for some fixed pre-specified $\epsilon>0$. Then, we have the following:

{\bf (a)} if $\rho(\M w)$ is the squared error prox function (see Section~\ref{sm-sq-err-prox1}), then
$\tilde{\phi}(\M  x ; \tau)$ has a Lipschitz continuous gradient with  
$\ML = \lambda_{\max}( A^\top A) (m - 1)/(m\epsilon)$.


{\bf (b)} If $\rho(\M w)$ is the entropy prox function (see Section~\ref{sm-ent-err-prox1}) then
$\tilde{\phi}(\M  x ; \tau)$ has a Lipschitz continuous gradient with  
$\ML = {\left( \max\limits_{i,j} | A_{ij} | \right)^2 \log m}/{\epsilon}.$ 
  \end{thm}

\begin{rem}
Note that there is a trade-off between the smoothness of the function $\tilde{\phi}(\M{x} ; \tau)$ and its accuracy in approximating $\hat{\phi}(\M{x})$. 
Its smoothness, measured by the Lipschitz constant of its gradient is $O(\frac{1}{\epsilon})$
and the approximation error is $O(\epsilon)$.
\end{rem}

\subsection{Smoothing for scheme~\eqref{eq:CanonLSE}}\label{sec:SmLSE2}
We will now consider smoothing schemes for the interpolant which is given by the linear program (LP) in \eqref{eq:CanonLSE}
with optimization variables $\alpha_{1}, \ldots, \alpha_{n}$.  Observe that the smoothing scheme described above in~\eqref{pwise-affine-gen1} works for  convex functions that admit a \emph{max-like} representation. The function~\eqref{eq:CanonLSE}, on the other hand, admits a \emph{min-like} representation, hence the smoothing scheme of Section~\ref{sec:SmLSE} does not directly apply. To circumvent this we consider the dual representation of the LP appearing in~\eqref{eq:CanonLSE}, which is given by the following LP:
\begin{equation}\label{eq:CanonLSE-dual}
\begin{aligned}
\check{\phi} (\M{x}) &\; = & \max \;  [ -\mu - \langle \B \zeta , \M{x} \rangle ]&
                                    &\;\;s.t.\;\; &    \hat\theta_{i} + \mu +  \langle \B \zeta , \M{X}_{i} \rangle \geq 0,& \;\;\; i = 1, \ldots, n, 
\end{aligned}
\end{equation}
where the optimization variables are $\mu \in \Re , \B \zeta \in \Re^{d}$. 
Consider a modified version of the estimator~\eqref{eq:CanonLSE-dual} obtained by adding a strongly convex 
prox function to the objective in problem~\eqref{eq:CanonLSE-dual}. This leads to the following convex program:
\begin{equation}\label{eq:CanonLSE-dual-sm1}
\begin{aligned}
\check \phi^{\text{Sm}} (\M{x}) =\max \left[ -\mu - \langle \B \zeta , \M{x} \rangle  - \tau \rho( \mu, \B \zeta) \right] \;\;
                                    s.t. \;\;    \hat\theta_{i} + \mu +  \langle\B  \zeta , \M{X}_{i} \rangle \geq 0, \quad  i = 1, \ldots, n,& 
\end{aligned}
\end{equation}
with variables $\mu \in \Re , \B \zeta \in \Re^{d}$.
For simplicity, consider the Euclidean prox function, i.e., 
$ \rho( \mu, \B \zeta ) = \half \left ( \mu^2 + \|\B \zeta \|_2^2 \right) $
with the respective spaces endowed with the standard Euclidean $\|\cdot\|_{2}$-norm. It follows from Theorem~\ref{main-thm-nest-1} that the function $\check \phi^{\text{Sm}}(\M{x})$ is differentiable in $\M{x}$ with Lipschitz continuous gradient $-\hat{\B \zeta}^{\tau}$, where $(\hat{\mu}^{\tau}, \hat{\B \zeta}^{\tau})$ is the unique optimal solution to problem~\eqref{eq:CanonLSE-dual-sm1}.

\subsection{Bias correction}\label{sec:BiasCorr}
Although~\eqref{eq:UnifBd1} and~\eqref{eq:UnifBd2} provide a two-sided bound for the maximal deviation between $\hat \phi$ (or $\check \phi$) and the smoothed estimator $\tilde \phi$ (or $\check \phi^{\text{Sm}}$),  Lemma~\ref{lem-unif-bound1} (see~\eqref{eq:Bound}) shows that the smoothed estimator is always less than $\hat \phi$. This bias of the smoothed estimator can be easily corrected for. It can be easily shown that the fitted $\hat{\B \theta} $ always has the same average as the sample mean (see e.g.,~\citet[Lemma 2.4]{SS11}), i.e., $\sum_{i=1}^n \hat \theta_i = \sum_{i=1}^n Y_i$. Let $\tilde{\B \theta} = (\tilde \theta_1,\ldots, \tilde \theta_n)$ denote the vector of values of the smoothed estimator at the $\M X_i$'s. It is natural to enforce that the bias corrected smoothed estimator should also have the same mean as the sample mean. This leads to the bias corrected smoothed estimator defined as 
	$\tilde \phi_{\text{BC}}(\M x) := \tilde \phi(\M x) + \frac{1}{n} \sum_{i=1}^n (\hat \theta_i -  \tilde \theta_i).$
We can similarly define the bias corrected smoothed estimator obtained using $\check \phi^{\text{Sm}}$. Figure~\ref{fig:SmoothPlots} shows the bias corrected estimator along with the LSE when $n =100$.   


\section{Lipschitz Convex Regression}\label{sec:unif-Lip-bound1}
The LSE described in \eqref{eq:CvxLSE} suffers from over-fitting, especially near the boundary of the convex hull of the design points $\M X_i$'s. The norms of the fitted subgradients $\hat{\B \xi_i}$'s near the boundary 
can become arbitrarily large as the sample size grows and there can be a large proportion of data points near the boundary, for $d > 1$. This, in turn, can deteriorate the overall performance of the LSE. 
In fact, even when $d=1$ it is expected that the convex LSE will be inconsistent at the boundary; see~\cite{B07} for a proof in the context of density estimation.

As a remedy to this over-fitting problem we propose LS minimization over the class of convex functions that are uniformly Lipschitz with a known bound. For a convex function $\psi: \mathfrak{X} \rightarrow \Re$, let us denote by $\partial \psi(\M x)$ the subdifferential (set of all subgradients) at $\M x \in \mathfrak{X}$, and by $\|\partial \psi(\M x)\|$ the maximum $\|\cdot\|_2$-norm of vectors in $\partial \psi(\M x)$, i.e., $\|\partial \psi(\M x)\| := \sup_{\B \zeta \in \partial \psi(\M x)} \|\B \zeta\|_2$. For $L >0$, consider the class $\C_L$ of convex functions with Lipschitz norm bounded by $L$, i.e.,
\begin{equation}\label{eq:C_L}
\C_L := \left\{\psi: \mathfrak{X} \rightarrow \Re \; |\ \psi \mbox{ is convex},\ \sup_{\M x \in \mathfrak{X}} \|\partial \psi(\M x) \| \le L\right \}.
\end{equation} 
Let $\hat \phi_{n,L}$ denote the LSE when minimizing the sum of squared errors over the class $\C_L$, i.e., 
\begin{equation}\label{eq:LipLSE}
\hat \phi_{n,L} \in \argmin_{\psi \in \C_L} \sum_{i=1}^n (Y_i - \psi(\M{X}_i))^2.
\end{equation} 
The above problem is an infinite dimensional optimization problem. 
Fortunately, as before, the solution to the above problem can be obtained by solving the following finite dimensional convex optimization problem:
\begin{equation}\label{eq:CvxLipLSE}
\begin{aligned}
\mini_{\B\xi_1, \ldots, \B\xi_n;  \B\theta} & \;\;\;\; \frac{1}{2}\| \M{Y} -  \B{\theta} \|_2^2  \\
s.t.& \;\;  \theta_{j} + \langle \B{\Delta}_{ij}, \B{\xi}_{j} \rangle \leq \theta_{i}; \; i,j = 1, \ldots, n; \\
&\;\;  \|\B \xi_j\|_2 \le L, \;\;\; j = 1, \ldots, n.
\end{aligned}
\end{equation}
To see that Problems~\eqref{eq:LipLSE} and~\eqref{eq:CvxLipLSE} are equivalent, it suffices to consider a solution to 
Problem~\eqref{eq:CvxLipLSE} and extend it to a member of the set $\C_L$  using~\eqref{eq:MaxLSE}.
Such an extension does not change the loss function and satisfies the feasibility condition of both problems~\eqref{eq:LipLSE} and~\eqref{eq:CvxLipLSE}.



We present the following result (proved in Section~\ref{proof:RateOfConv}) concerning the asymptotic rate of convergence of the convex Lipschitz LSE (i.e., Problem~\eqref{eq:CvxLipLSE}). 
\begin{thm}\label{thm:RateOfConv}
Consider observations $\{(\M X_i, Y_i): i = 1,\ldots, n\}$ such that
$	Y_i = \phi(\M{X}_i) + \epsilon_i,$
where $\phi:\Re^d \rightarrow \Re$ is an unknown convex function. We assume that (i) the support of $\M X$ is $\mathfrak{X} =[0,1]^d$; (ii) $\phi \in \C_{L_0}$ for some $L_0 >0$; (iii) the $\M X_i \in \mathfrak{X}$'s are fixed constants; and (iv) $\epsilon_i$'s are independent mean zero sub-Gaussian errors (i.e., there exists $\sigma^2 >0$ such that for every $t \in \Re$ one has $\E [e^{t \eps_1} ]\le e^{\sigma^2 t^2/2}$). Given data from such a model, we have for any $L>L_0$,
\begin{equation}\label{eq:L_2Rate}
	\frac{1}{n} \sum_{i=1}^n (\hat \phi_{n,L}(\B X_i) - \phi(\B X_i))^2 = O_\pr(r_n),
\end{equation}
where      
\begin{equation}\label{eq:Rates}
r_n = \left\{ \begin{array}{ll}
          n^{-2/(d+4)} & \mbox{if $d = 1,2,3$},\\
        n^{-1/4} (\log n)^{1/2}  & \mbox{if $d=4$},\\
        n^{-1/d} & \mbox{if $d\ge 5$}.\end{array} \right.
\end{equation} 
\end{thm}
\begin{rem}
The above result follows from known metric entropy bounds on the class of all convex functions that are uniformly bounded and uniformly Lipschitz (under the uniform metric) and the theory on the rates of convergence of LSEs; see e.g.,~\citet[Theorem 9.1]{vdG00}. Note that the class of all convex functions is much larger (in fact, the set is not totally bounded) and thus finding the rate of convergence of the convex LSE (see~\eqref{eq:MaxLSE}) is a much harder problem; in fact it is still an open problem. 
\end{rem}
\begin{rem}
The assumption that $\mathfrak{X} =[0,1]^d$ can be extended to any compact subset of $\Re^d$.
\end{rem}
\begin{rem}
Recently, during the preparation of the manuscript, we became aware of the work of~\cite{B15} and~\cite{L14}, where LSEs obtained over different sub-classes of convex functions are proposed and studied.~\cite{B15} propose Lipschitz convex regression (where instead of the $\|\cdot\|_2$-norm on the subgradients, the authors study the $\|\cdot\|_\infty$-norm) with an additional boundedness constraint and study the rates of convergence of the obtained LSE. \cite{L14} studies $\|\cdot\|_\infty$-Lipschitz convex regression and the main result in the paper is similar to Theorem~\ref{thm:RateOfConv}. We, of course, deal with a different formulation as we constrain the $\|\cdot\|_2$-norm of the subgradients. Moreover, the proof of Theorem~\ref{thm:RateOfConv} can be mimicked to give a shorter and simpler proof of the main result in~\cite{L14}. 
\end{rem}

The following lemma (proved in Section~\ref{proof-lem:sm-approx0}) shows that  the estimator obtained upon applying the smoothing method to the estimator obtained from Problem~\eqref{eq:CvxLipLSE} lies in $\C_L$.
\begin{lem}\label{lem:sm-approx0}
Let $\tilde{\phi}(\M x; \tau)$ be the smoothed estimator obtained via the scheme in Section~\ref{sec:SmLSE} for estimator~\eqref{pwise-affine-gen1} with $\hat{\theta}_{i}, \hat{\B\xi}_{i}$'s obtained as solutions to Problem~\eqref{eq:CvxLipLSE}. Then $\| \nabla \tilde{\phi}(\M x; \tau) \|_{2} \leq L$.
\end{lem}

\subsection{Algorithm for Problem~\eqref{eq:CvxLipLSE}}\label{sec:algo-unif-Lip1}

The convex optimization problem~\eqref{eq:CvxLipLSE} can be solved by interior point methods. However, as in the case of Problem~\eqref{eq:CvxLSE} off-the-shelf interior point methods
have difficulty scaling to large $n$. In this vein, we propose a simple variant of the ADMM algorithmic framework described for Problem~\eqref{eq:CvxLSE} that can be applied to solve Problem~\eqref{eq:CvxLipLSE}.
To see this we consider the following equivalent representation of Problem~\eqref{eq:CvxLipLSE}:
\begin{equation}\label{eq:CvxLipLSE-admm1}
\begin{aligned}
\mini_{\B\xi_1, \ldots, \B\xi_n;  \B\theta; \B{\eta}} & \;\;  \;\; \frac{1}{2}\| \M{Y} -  \B{\theta} \|_2^2\\
s.t. \;\;\; &\;\;\;  \eta_{ij} =  \theta_{j} + \langle \B{\Delta}_{ij}, \B{\xi}_{j} \rangle -  \theta_{i};   \\
& \;\;\; \eta_{ij} \leq 0 ; \; i = 1, \ldots, n, \; j = 1, \ldots, n;\\ & \|\B \xi_j\|_2 \le L, \;\;\; j = 1, \ldots, n;
\end{aligned}
\end{equation}
and the corresponding augmented Lagrangian similar to~\eqref{eq:CvxLSE-admm1}.
We use an ADMM algorithm with a similar block splitting strategy as in Algorithm~\ref{algo1-admm} --- the main difference with Algorithm~\ref{algo1-admm} being update~\eqref{admm-updxi1} where, we need to solve:
\smallskip \\
Step 1.~Update the subgradients $(\B\xi_1, \ldots, \B\xi_n)$:
\begin{equation}\label{admm-updxi3}
(\B\xi^{(k+1)}_1, \ldots, \B\xi^{(k+1)}_n) \in \argmin_{\B\xi_1, \ldots, \B\xi_n :  \| \B\xi_{j}\|_{2} \leq L, \forall j } {\mathcal L}_\rho((\B\xi_1, \ldots, \B\xi_n;   \B\theta^{(k)}; \B{\eta}^{(k)}); \B\nu^{(k)}).
\end{equation} 
The above problem is a QP in $nd$ variables with $nd$ constraints. Note that the optimization problem is separable in the variables 
$\B\xi_{j}$'s, for $j = 1, \ldots, n$. Thus solving Problem~\eqref{admm-updxi3} is equivalent to solving  for each $j$ the following constrained variant of the 
convex problem~\eqref{admm-updxi1-simp}:
\begin{equation}\label{socp-ls-1}
\mini_{\B\xi_{j}}\;\; \sum_{i=1}^{n} \left( \bar{\eta}_{ij} - \langle \Delta_{ij}, \B\xi_{j} \rangle \right)^2 \;\;\;s.t.\;\;\; \|\B\xi_{j} \|_{2} \leq L.
\end{equation}
The above problem, unlike Problem~\eqref{admm-updxi1}, does not admit a closed form solution. However, Problem~\eqref{socp-ls-1} is a Second Order Cone Program (SOCP) with $d$ variables
and since $d$ is typically quite small (e.g., 10 or 20), solving it is quite cheap. Problem~\eqref{socp-ls-1} can thus 
be solved by using an off-the-shelf interior point method. It can also be solved via specialized algorithms --- we develop our own algorithm for Problem~\eqref{socp-ls-1} that we describe in Section~\ref{sec:append-socp-1}.

\begin{figure}[h!]
\begin{center}
\resizebox{\textwidth}{0.15\textheight}{\begin{tabular}{cc}
\sf \scriptsize {Risk} &\sf \scriptsize{Training Error}  \\
\includegraphics[height=2.2in,width= 2.85in,angle= 0,trim =1.cm 1.6cm 1.cm 2cm, clip = true]{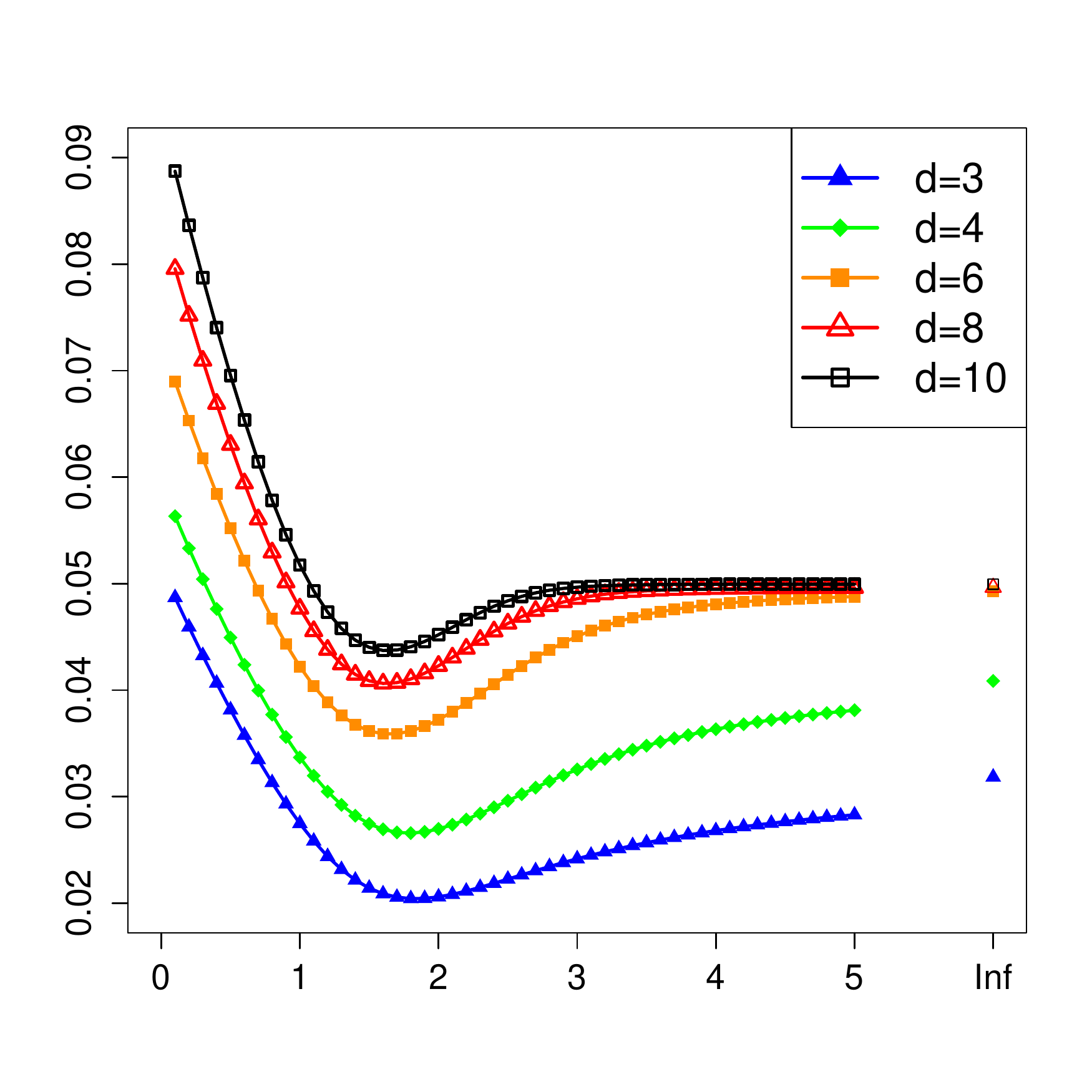} \hspace{3mm} &
\includegraphics[height=2.2in,width= 2.85in,angle= 0,trim =1.cm 1.6cm 1.cm 2cm, clip = true]{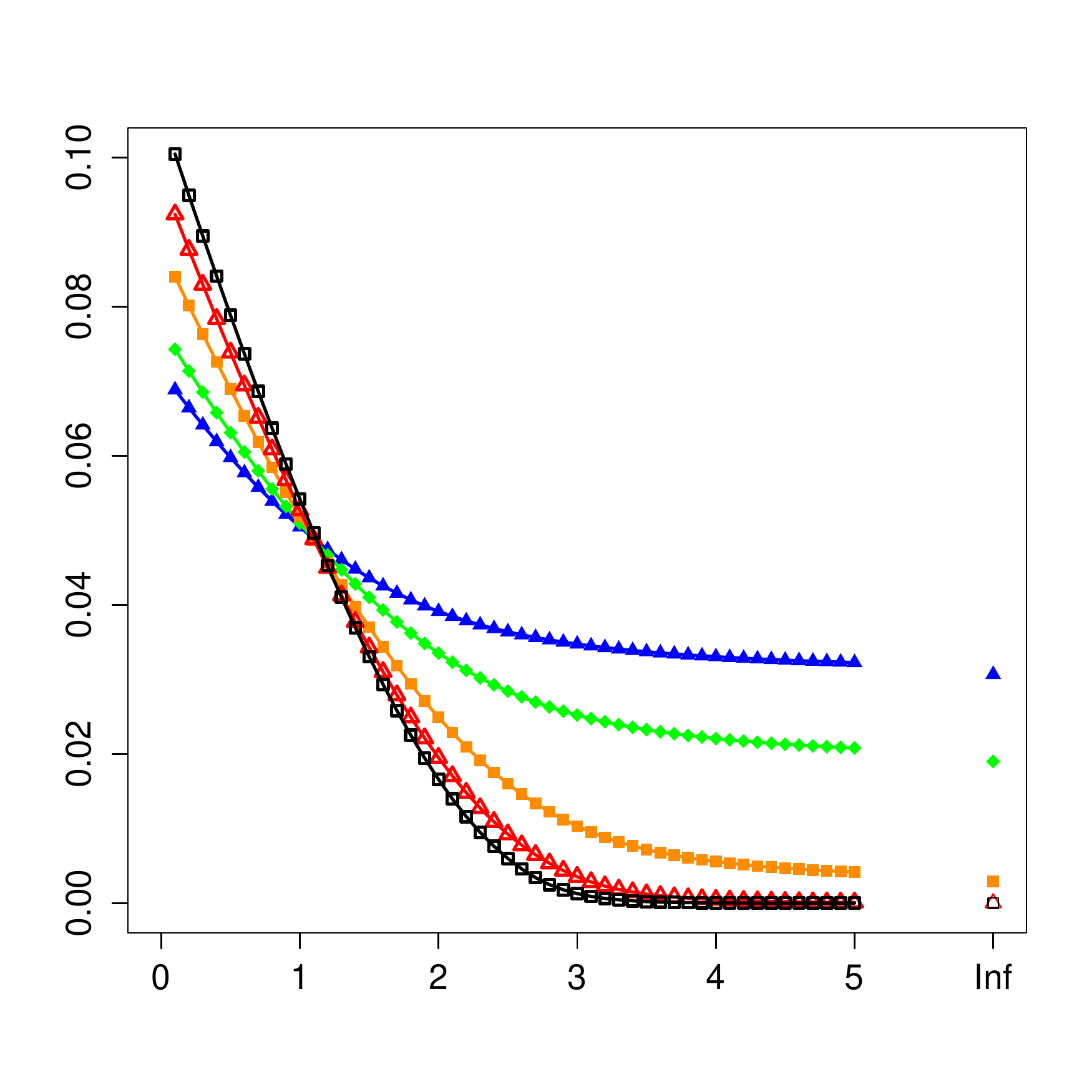}\\
\sf \scriptsize {Lipschitz bound ($L$)} &  \sf \scriptsize{Lipschitz bound ($L$)}  \\
\end{tabular}}
\caption{{\small{[Left panel]: the simulated risk of the Lipschitz convex estimator as the Lipschitz bound $L$ varies (L = Inf gives the usual convex LSE) for 5 different dimension values ($d$). [Right panel]: the training error as the Lipschitz bound $L$ varies, for the same examples appearing in the left panel.}}}
\label{fig:LipError}
\end{center}
\end{figure}

In Figure~\ref{fig:LipError} we show the performance of Lipschitz convex regression when $n=100$, $\phi(\M x) = \|\M x\|_2^2$ and $d$ varies in $\{3, 4, 6, 8,10\}$. The left panel illustrates that a proper choice of $L$ can lead to substantial reduction in the risk of the estimator (as measured by $\E_\phi \sum_{i=1}^n (\hat \phi_{n,L}(\M X_i) - \phi(\M X_i))^2$) in estimating $\phi$. The right panel shows that as the dimension $d$ grows, the usual convex LSE overfits the data (for $d=10$ the training error is essentially 0).

To choose an optimal tuning parameter $L$, the Lipschitz constant that minimizes the risk, we advocate the use of cross-validation; see~e.g.,~\citet{ELS09}. In our extensive simulation studies we observed that tenfold cross-validation  works quite well. We use the ``one-standard error'' rule in cross-validation where we choose the smallest $L$ whose error is no more than one standard error above the error of the best Lipschitz parameter.

\section{Convex Functions with Coordinate-wise Monotonicity}\label{multi-cwise-mono-1}
In this section we consider a variation of the problem proposed in~\eqref{eq:LSE}, where the underlying convex regression function, restricted to each of its coordinates, is assumed to be monotone (i.e., increasing or decreasing). Estimation of such functions have a wide range of applications, especially in demand and production frontiers in economics; see e.g.,~\citet{V82},~\citet{V84},~\citet{M94},~\citet{Y98},~\citet{K08} and the references therein. 

To fix ideas, let us assume that the function to be estimated is known to be convex and non-decreasing coordinate-wise and we have data $\{(\M X_i, Y_i): i = 1,\ldots, n\}$. This leads to the following infinite dimensional optimization problem:
\begin{equation}\label{conv-increase-1}
\begin{aligned}
\mini_{\psi} \;\; & \sum_{i=1}^n (Y_i - \psi(\M{X}_i))^2 
\end{aligned}
\end{equation}
where the minimization is carried over all convex functions $\psi$ that are non-decreasing in each of its coordinates, i.e.,  $\nabla_{k} \psi(\M{x}) \geq 0$ for all $\M{x}$ and 
for all $k= 1, \ldots, d$ (here $\nabla_{k} \psi(\M{x})$ denotes the $k$'th coordinate of a subgradient of $\psi$ at $\M{x}$). Not surprisingly, this seemingly infinite dimensional optimization problem can also be cast as the following (finite dimensional) QP:
\begin{equation}\label{conv-reg1}
\begin{aligned}
\mini_{\B\xi_1, \ldots, \B\xi_n;  \B\theta} & \;\;\;\; \frac{1}{2}\| \mathbf{Y} -  \B{\theta} \|_2^2\\
s.t. \;\;\; &\;\;\;  \theta_{j} + \langle \B{\Delta}_{ij}, \B{\xi}_{j} \rangle \leq \theta_{i}; \;\;\; i = 1, \ldots, n, \; j = 1, \ldots, n,\\
 \;\;\; &\;\;\;  {\xi}_{ji}\geq 0; \; j = 1, \ldots, n, \; i = 1, \ldots, d, 
\end{aligned}
\end{equation}
where the notations used above are the same as in Problem~\eqref{eq:CvxLSE} with $\xi_{ji}, i = 1, \ldots, d,$ being the coordinates of the subgradient $\B\xi_{j}$. Note that the constraints ${\xi}_{ji}\geq 0$, for $j = 1, \ldots, n$, represent that the $i$'th coordinate of the subgradients of the function evaluated at the data points $\M{X}_{j}$'s are non-negative. This is equivalent to the function restricted to the $i$'th coordinate being non-decreasing. The above formulation resembles Problem~\eqref{eq:CvxLSE} with the exception of the additional $nd$ constraints ${\xi}_{ji} \geq 0$ for all $i,j$.

To see why Problems~\eqref{conv-reg1} and~\eqref{conv-increase-1} are equivalent, observe that any solution $\{\B\xi^*_i\}_{i=1}^n$ and $\B\theta^*$ of Problem~\eqref{conv-reg1} 
can be extended to a convex function on $\Re^d$ by the rule~\eqref{eq:MaxLSE}. Note that $\hat{\phi}_n$ thus defined is convex in $\Re^d$.  Any subgradient of the function $\hat{\phi}_n$ when restricted to the $k$'th coordinate is non-negative and hence the function $\hat{\phi}_n$ is non-decreasing in each coordinate. Furthermore, this function has the same loss function as the optimal objective value of Problem~\eqref{conv-reg1}.
Thus solving Problem~\eqref{conv-reg1} is equivalent to solving Problem~\eqref{conv-increase-1}.


We use an ADMM type algorithm to solve \eqref{conv-reg1} in a similar manner as \eqref{eq:CvxLSE-admm1}, by considering the following equivalent representation for~\eqref{conv-reg1}:
\begin{equation}\label{conv-reg1-admm1}
\begin{aligned}
\mini_{\B\xi_1, \ldots, \B\xi_n;  \B\theta; \B{\eta}} & \;\;\;\; \frac{1}{2}\| \mathbf{Y} -  \B{\theta} \|_2^2\\
s.t. \;\;\; &\;\;\;   \eta_{ij} =  \theta_{j} + \langle \B{\Delta}_{ij}, \B{\xi}_{j} \rangle -  \theta_{i};   \\
& \;\;\; \eta_{ij} \leq 0 ; \; i = 1, \ldots, n; \; j =1, \ldots, n, \\
& \;\;\; {\xi}_{ji}\geq 0; \; j = 1, \ldots, n; \; i = 1, \ldots, d, 
\end{aligned}
\end{equation}
where $\B\eta=((\eta_{ij})) \in \Re^{n \times n}$ is a matrix with $(i,j)$'th entry $\eta_{ij}$. 
We then consider the augmented Lagrangian corresponding to the above formulation, similar to~\eqref{Lag}, and employ a multiple block version of ADMM along the lines of Algorithm~\ref{algo1-admm}. The main difference with Algorithm~\ref{algo1-admm} is the update~\eqref{admm-updxi1} --- we now need to consider the following constrained QP:
\smallskip \\
Step 1.~Update the subgradients $(\B\xi_1, \ldots, \B\xi_n)$:
\begin{equation}\label{admm-updxi2}
(\B\xi^{(k+1)}_1, \ldots, \B\xi^{(k+1)}_n) \in \argmin_{\B\xi_1, \ldots, \B\xi_n : \B\xi_{ji} \geq 0 ,  \; \forall i, j } {\mathcal L}_\rho((\B\xi_1, \ldots, \B\xi_n;   \B\theta^{(k)}; \B{\eta}^{(k)}); \B\nu^{(k)}).
\end{equation} 
The above problem is a QP in $nd$ variables with $nd$ constraints. Note that the optimization problem is separable in the variables $\B\xi_{j}$'s, for $j = 1, \ldots, n$. Thus solving Problem~\eqref{admm-updxi2} is equivalent to solving  for each $j$ the following non-negative LS  problem:
\begin{equation}\label{nng-ls-1}
\begin{aligned}
 \mini_{\B\xi_{j}}~& {\small  \B\xi^\top_{j}\left( \sum_{i=1}^n  \B\Delta_{ij}\B{\Delta}^\top_{ij} \right) \B\xi_{j} -  2 \left\langle  \sum_{i=1}^{n} \left ( \frac{1}{\rho} \nu_{ij} + \eta_{ij}  -  (\theta_{j} -   \theta_{i}) \right) \B\Delta_{ij} , \B\xi_{j} \right\rangle} \\
s.t.~&~~ ~~~~\xi_{jk} \geq 0 , \;\;\;k = 1, \ldots, d . 
 \end{aligned}
 \end{equation}
Unlike Problem~\eqref{admm-updxi1}, Problem~\eqref{nng-ls-1} does not admit a closed form solution. However, Problem~\eqref{nng-ls-1} is a QP with $d$ variables and since $d$ is typically small, solving it is quite cheap. Problem~\eqref{nng-ls-1} can be solved by using an off-the-shelf interior point method for each $j$. We however use our own implementation of one-at-a-time coordinate descent for the above optimization problem. The procedure is described in Section~\ref{appendix-cd-1}.

\begin{rem}
Note that the framework described above can easily accommodate other variants of~\eqref{conv-reg1}. For example, if the convex function restricted to each of the coordinate directions is assumed to be non-increasing then one needs to use the constraints $\xi_{ij} \leq 0$ for all $i,j$ --- which can be addressed by using a minor variant of Problem~\eqref{nng-ls-1}. Similarly, it is also possible to accommodate the case where the function is assumed to be non-decreasing in some of the coordinates and non-increasing in some others. 
\end{rem}

The following lemma (proved in Section~\ref{proof:lem:sm-approx1}) shows that our smoothing operation does not change the sign of the coordinate-wise subgradients of the function.
\begin{lem}\label{lem:sm-approx1}
Let $\tilde{\phi}(\M x; \tau)$ be the smoothed estimator obtained via the scheme in Section~\ref{sec:SmLSE} for the estimator~\eqref{pwise-affine-gen1}, with $\hat{\theta}_{i}, \hat{\B\xi}_{i}$'s obtained as solutions to Problem~\eqref{conv-reg1}. Then $\nabla_{k} \tilde{\phi}(\M x; \tau) \geq 0$, for $k=1,\ldots, d$, i.e., the smooth estimator is also coordinate-wise increasing.
\end{lem}

\section{Numerical Experiments}\label{sec:NumExp}

\begin{figure}[h!]
\centering
\resizebox{\textwidth}{0.2\textheight}{\begin{tabular}{ l c  cc}
& \scriptsize{\sf {Example 1  (n=500,d=2)}}& \scriptsize{\sf {Example 2   (n=1000,d=10)}} & \scriptsize{\sf{Example 6 (n=3449,d=4)}}\\
\rotatebox{90}{\sf {\scriptsize{~~~~~~~~$\log_{10}$(Primal Feasibility)}}}&
\includegraphics[width=0.32\textwidth,height=0.22\textheight,  trim =1.0cm 2.5cm 1.cm 1.5cm, clip = true ]{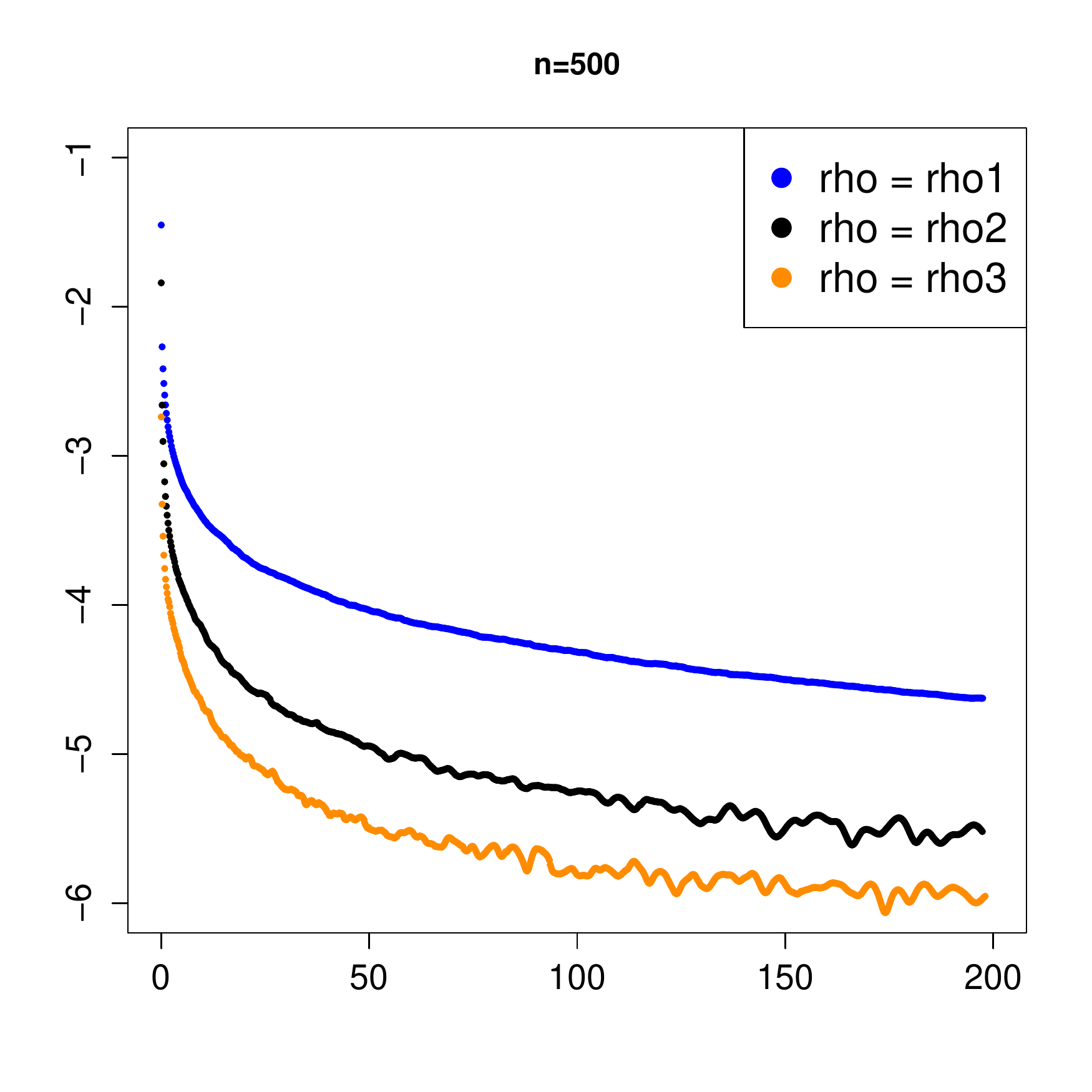}&
\includegraphics[width=0.32\textwidth,height=0.22\textheight,  trim =2.0cm 2.5cm 1.cm 1.5cm, clip = true ]{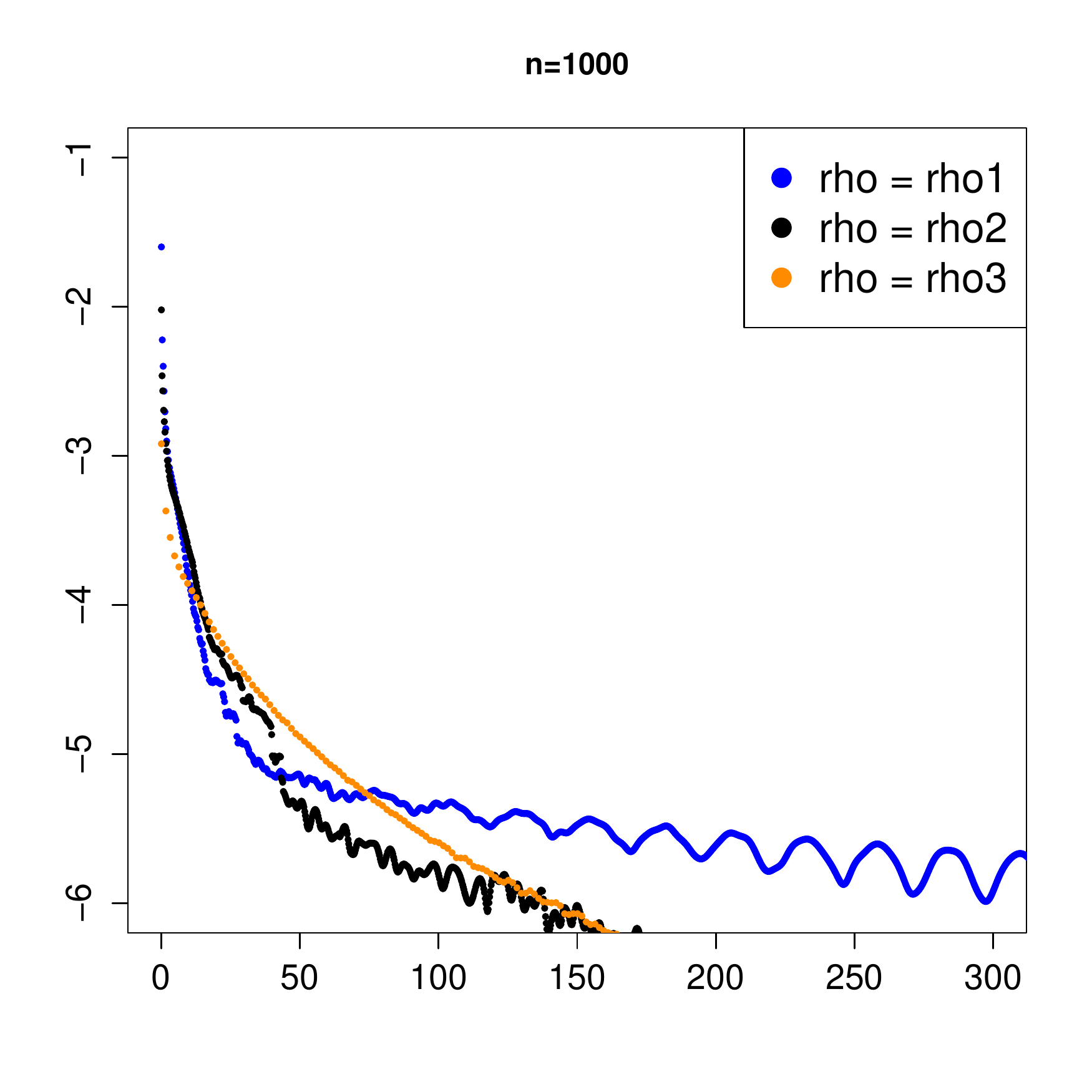}&
\includegraphics[width=0.32\textwidth,height=0.22\textheight,  trim =2.0cm 2.5cm 1.cm 1.5cm, clip = true ]{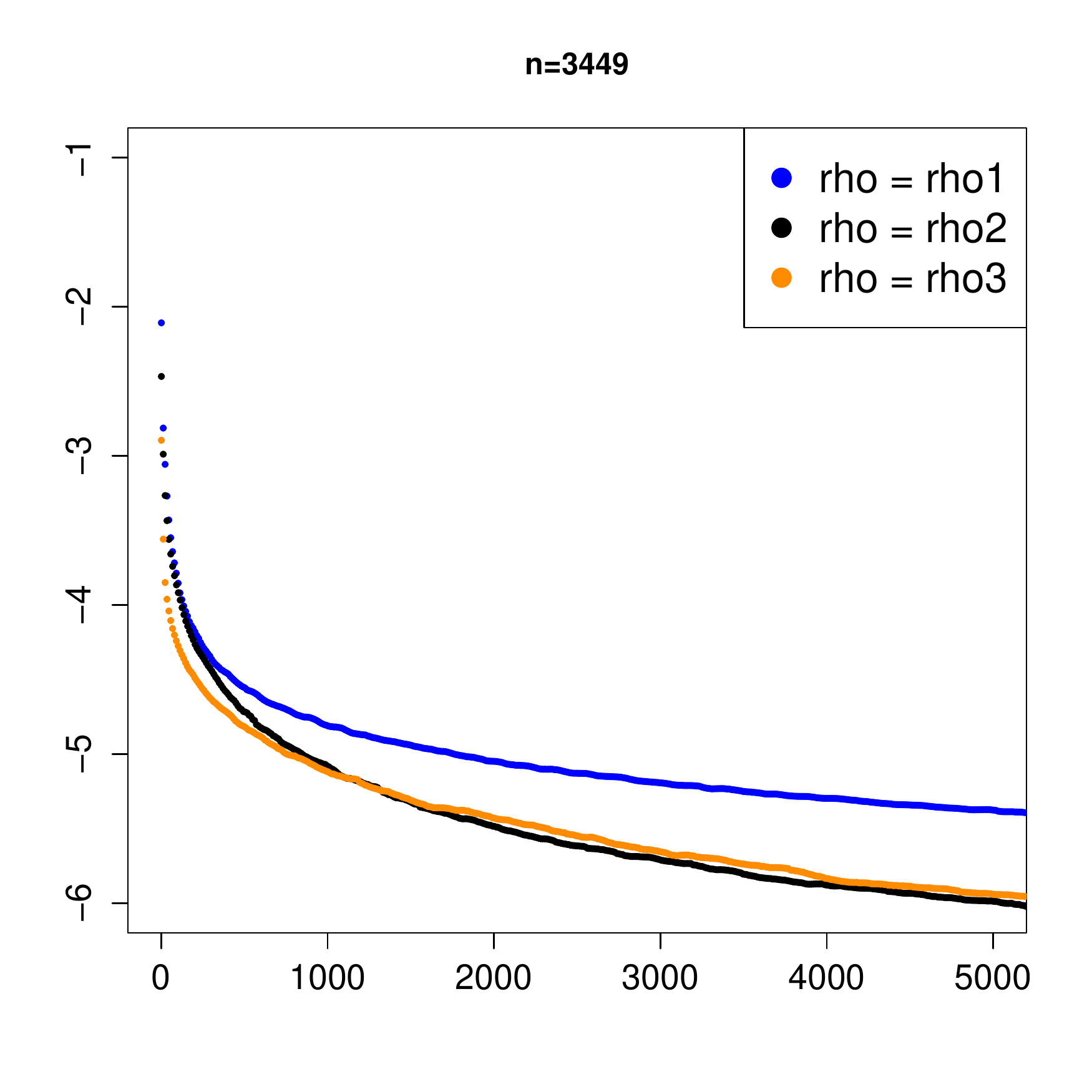} \medskip \\
\rotatebox{90}{\sf {\scriptsize{~~~~~~~~~~~~~$\log_{10}$(Gradient)}}}&
\includegraphics[width=0.32\textwidth,height=0.22\textheight,  trim =1.0cm 1.2cm 1.cm 1.5cm, clip = true ]{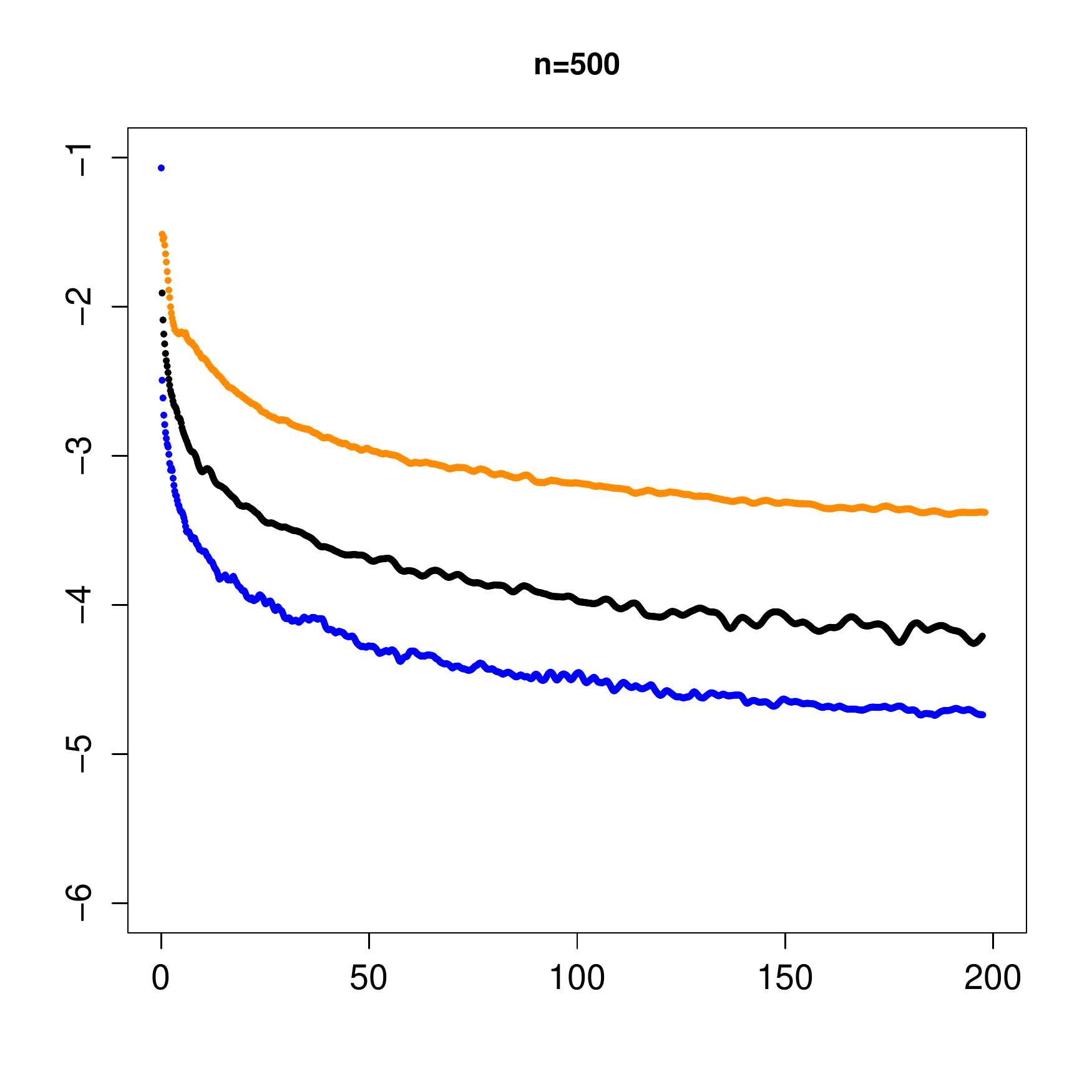}&
\includegraphics[width=0.32\textwidth,height=0.22\textheight,  trim =2.0cm 1.2cm 1.cm 1.5cm, clip = true ]{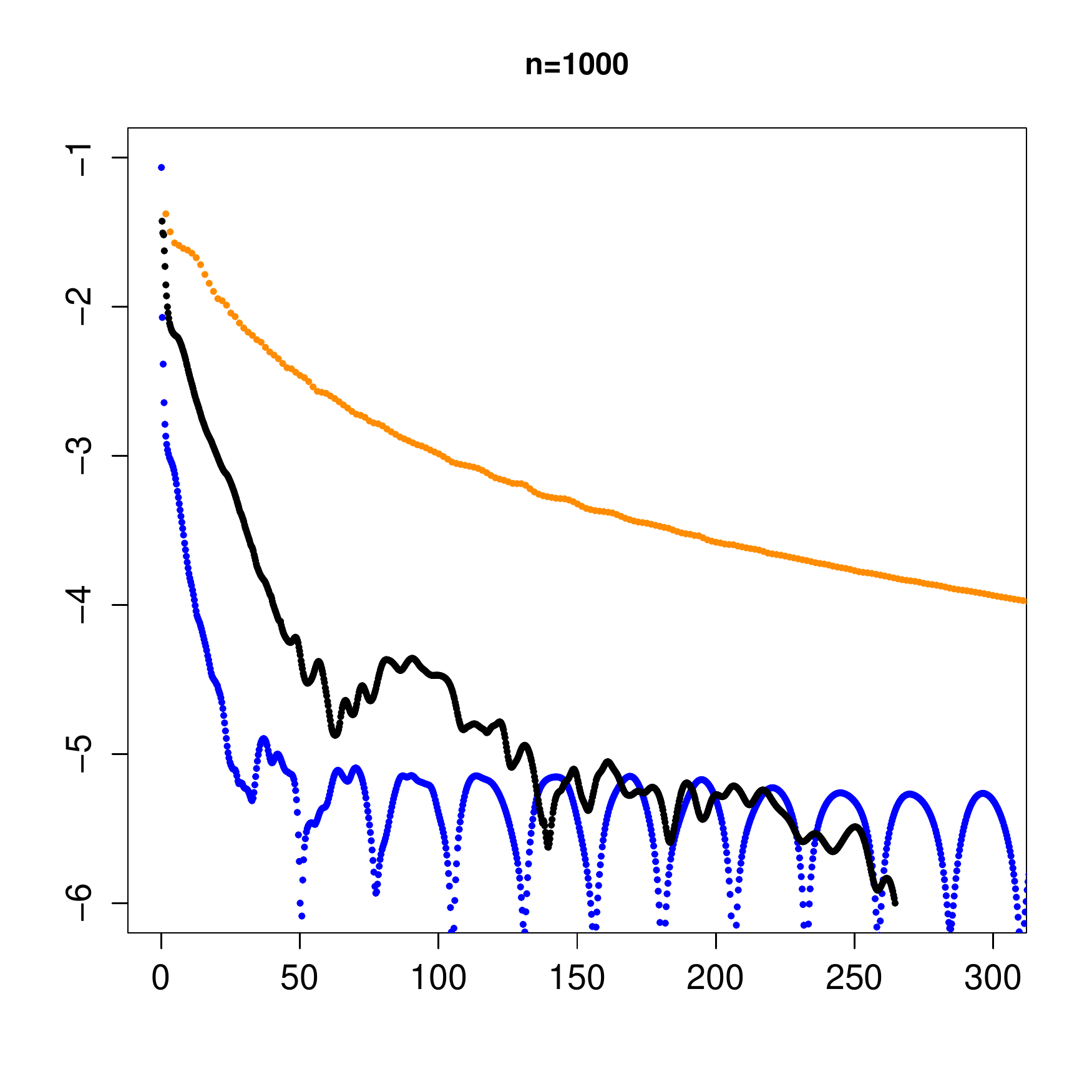}&
\includegraphics[width=0.32\textwidth,height=0.22\textheight,  trim =2.0cm 1.2cm 1.cm 1.5cm, clip = true ]{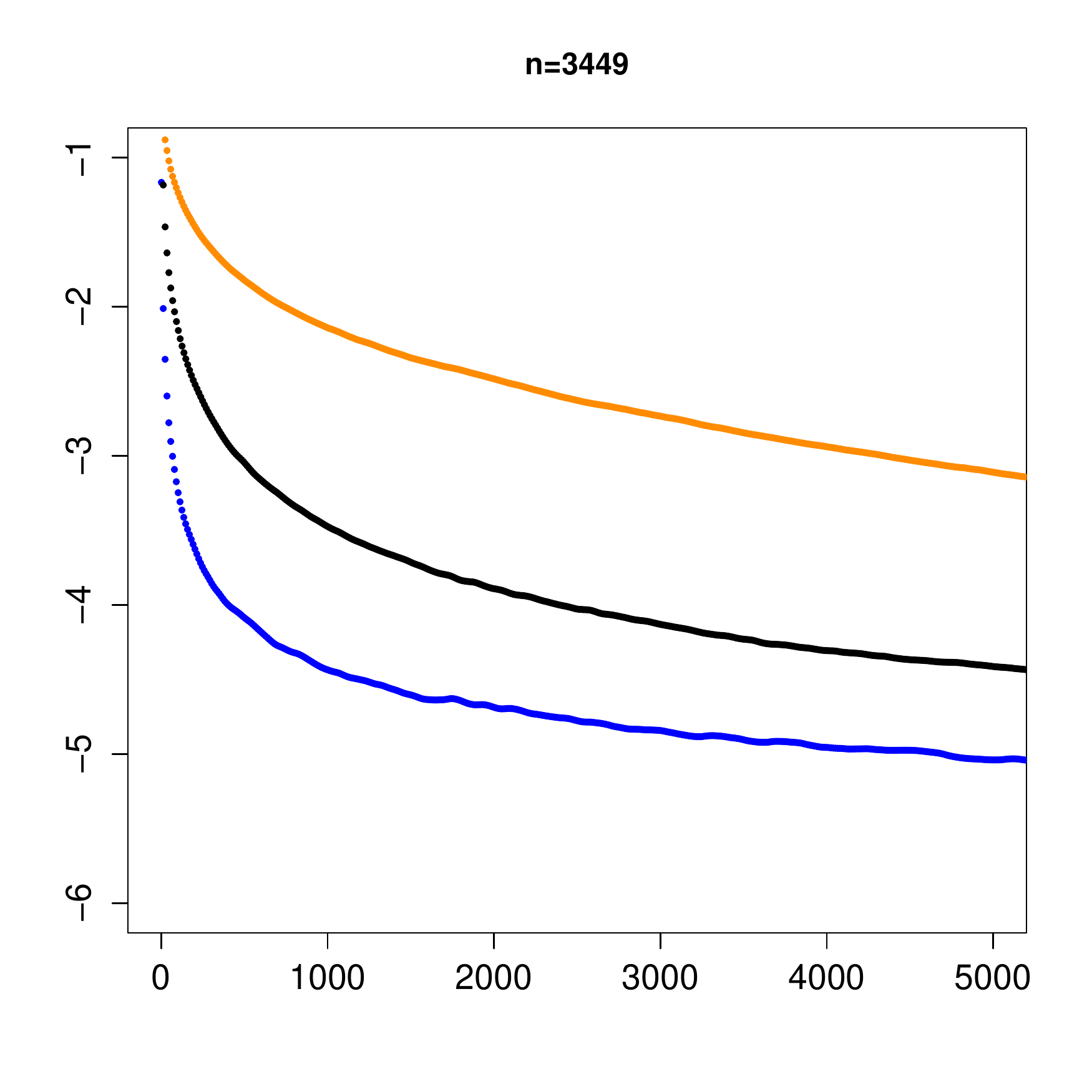} \\
& \scriptsize{\sf {Time  (in secs)}}& \scriptsize{\sf {Time  (in secs)}} & \scriptsize{\sf{Time  (in secs)}}\\
\end{tabular}}
\caption{{\small{The evolution of Algorithm~1 with time, for three different examples: the right panel is a real-data example, the remaining are synthetic. The top panel shows the primal feasibility convergence with time and bottom panel shows the evolution of the $\|\cdot\|_{2}$-norm of the gradient with respect to $\B{\theta}$.  Three different $\rho$ values, denoted by `rho1', `rho2', `rho3', were taken to be ${0.1}/{n}, {1}/{n},{10}/{n}$ respectively. 
 Note that all the algorithms were allowed to run for a long time and hence the solutions obtained have high accuracy $\approx 10^{-6}$. The figures show that lower accuracy solutions are obtained faster, see also, 
Table~\ref{tab:algo1-algo3}.}}}\label{fig:compute_times_fig1}
\end{figure}

\begin{table}[h!]
\centering
\resizebox{.95\textwidth}{0.3\textheight}{\begin{tabular}{cc}
\begin{tabular}{|cccc|}
\multicolumn{4}{c}{Example 2 ($n=1000,d=10$)} \\ 
  \hline
Primal Feas & Gradient & Algorithm~2 & Algorithm~1 \\ Error&Error &Time (secs)&Time (secs)\\  \hline
1e-03 & 1e-02 & 2.043 & 2.240 \\ 
  1e-03 & 1e-03 & 15.444 & 20.224 \\ 
  1e-03 & 1e-04 & 28.546 & 40.457 \\ 
  1e-04 & 1e-02 & 14.146 & 14.373 \\ 
  1e-04 & 1e-03 & 15.444 & 20.224 \\ 
  1e-04 & 1e-04 & 28.546 & 40.457 \\ 
   \hline
\end{tabular}&
\begin{tabular}{|cccc|}
\multicolumn{4}{c}{Example 3 ($n=473,d=4$)} \\ 
  \hline
Primal Feas & Gradient & Algorithm~2 & Algorithm~1 \\ Error&Error &Time (secs)&Time (secs)\\  \hline
1e-03 & 1e-02 & 0.523 & 0.535 \\ 
  1e-03 & 1e-03 & 3.413 & 3.537 \\ 
  1e-03 & 1e-04 & 17.034 & 17.290 \\ 
  1e-04 & 1e-02 & 1.870 & 1.935 \\ 
  1e-04 & 1e-03 & 3.413 & 3.537 \\ 
  1e-04 & 1e-04 & 17.034 & 17.290 \\ 
   \hline
\end{tabular}\\ 
&\\
\begin{tabular}{|cccc|}
\multicolumn{4}{c}{Example 4 ($n=558,d=3$)} \\ 
  \hline
Primal Feas & Gradient & Algorithm~2 & Algorithm~1 \\ Error&Error &Time (secs)&Time (secs)\\  \hline
1e-03 & 1e-02 & 0.988 & 0.976 \\ 
  1e-03 & 1e-03 & 16.116 & 16.074 \\ 
  1e-03 & 1e-04 & 143.181 & 103.888 \\ 
  1e-04 & 1e-02 & 7.295 & 7.282 \\ 
  1e-04 & 1e-03 & 16.116 & 16.074 \\ 
  1e-04 & 1e-04 & 156.612 & 103.888 \\ 
   \hline
\end{tabular}&
\begin{tabular}{|cccc|}
\multicolumn{4}{c}{Example 5 ($n=1000,d=5$)} \\ 
  \hline
Primal Feas & Gradient & Algorithm~2 & Algorithm~1 \\ Error&Error &Time (secs)&Time (secs)\\  \hline
1e-03 & 1e-02 & 2.474 & 2.486 \\ 
  1e-03 & 1e-03 & 62.338 & 62.493 \\ 
  1e-03 & 1e-04 & 1226.499 & 516.286 \\ 
  1e-04 & 1e-02 & 39.143 & 39.951 \\ 
  1e-04 & 1e-03 & 62.338 & 62.493 \\ 
  1e-04 & 1e-04 & 1226.499 & 516.286 \\ 
   \hline
\end{tabular}\\
&\\
\begin{tabular}{|cccc|}
\multicolumn{4}{c}{Example 6 ($n=3449,d=4$)} \\ 
  \hline
Primal Feas & Gradient & Algorithm~2 & Algorithm~1 \\ Error&Error &Time (secs)&Time (secs)\\  \hline
1e-03 & 1e-02 & 73.743 & 75.738 \\ 
  1e-03 & 1e-03 & 448.043 & 459.826 \\ 
  1e-03 & 1e-04 & 3445.433 & 2392.528 \\ 
  1e-04 & 1e-02 & 118.601 & 122.275 \\ 
  1e-04 & 1e-03 & 448.043 & 459.826 \\ 
  1e-04 & 1e-04 & 3445.433 & 2392.528 \\ 
   \hline
\end{tabular} &
\begin{tabular}{|cccc|}
\multicolumn{4}{c}{Example 7 ($n=1500,d=4$)} \\ 
  \hline
Primal Feas & Gradient & Algorithm~2 & Algorithm~1 \\ Error&Error &Time (secs)&Time (secs)\\  \hline
1e-03 & 1e-02 & 12.171 & 12.304 \\ 
  1e-03 & 1e-03 & 66.701 & 67.982 \\ 
  1e-03 & 1e-04 & 324.665 & 330.774 \\ 
  1e-04 & 1e-02 & 28.217 & 28.691 \\ 
  1e-04 & 1e-03 & 66.701 & 67.982 \\ 
  1e-04 & 1e-04 & 324.665 & 330.774 \\ 
   \hline
\end{tabular}
\end{tabular}}\caption{{\small{Table showing the times (in secs) taken for Algorithm~1 and Algorithm~2 to reach solutions of different accuracy levels of the primal feasibility gap and the norm of the gradient, as defined in the text. For Algorithm~1 we took the choice $\rho = {1}/{n}$. Observe that the performances of Algorithms~1 and 2 are quite similar in most instances.}}} \label{tab:algo1-algo3}
\end{table}

\subsubsection*{Description of datasets}

In all the synthetic examples below, we generate $n$ samples as $y_{i} = \mu_{i} + \epsilon_{i}, i = 1, \ldots, n$, where $\mu_{i} := \phi(\M{X}_{i})$ is the value of a function $\mu$ evaluated at the $i$'th datapoint $\M{X}_{i} \in \Re^d$. The errors $\epsilon_{i}$'s are assumed to be i.i.d.~$N(0, \sigma^2)$, for $i = 1, \ldots, n$.
We define the Signal to Noise  Ratio (SNR) as $\text{SNR}=\text{Var}(\mu)/\text{Var}(\epsilon)$. The different examples that are studied are~given~below.

\noindent {\bf Example 1:} Here, we took $n=500,d=2$ and the covariates were generated from a random Uniform ensemble with mean zero. We took
$\phi(x) = \|x\|_{2}^2$  and the value of $\sigma^2$ was adjusted so that SNR = 3. Both the features and the response were mean-centered and standardized to have unit $\|\cdot\|_{2}$-norm before being \emph{fed} into the solvers.

\noindent {\bf Example 2:}  This is similar to Example 1 with a larger problem size: $n=1000, d= 10$.

\noindent {\bf Example 3:} This is a real dataset with $n=473$ and $d=4$. The dataset, which appears in the paper~\cite{wang2013estimating}, was downloaded from the link~\url{http://www.nber.org/data/nbprod2005.html}. Here, we took the response as the total value of shipment. The four independent variables were: total real capital stock,  production worker hours,  number of non-production workers and number of production worker hours.  Based on exploratory data-analysis, we took a log-transform of each of the covariates. We mean-centered and scaled each of the covariates and the response, so that they have unit $\|\cdot\|_{2}$-norm.


\noindent {\bf Example 4:} In this example, $n = 558$ and $d =3$. This dataset was taken from the link~\url{http://www.econ.kuleuven.ac.be/GME} (see~\cite{verbeek2008guide}) and contains production data for 569 Belgian firms, from the year 1996. The response was taken as the negative logarithm of the value added by a worker. The predictor variables were: amount of capital (in terms of total fixed assets at the end of 1995), labour and wages\footnote{Upon exploratory analysis, we found some possible outliers in the covariate-space, which were discarded via some  simple pre-processing method.}.  As in the other cases, both the features and response were mean-centered and standardized to have unit $\|\cdot\|_{2}$-norm.

\noindent {\bf Example 5:}
In this synthetic example, we took $n=1000$ and $d=5$, with SNR=3. Here, the underlying convex function was taken as: 
$\phi(x) = (5x_{1} + 0.5x_{2} + x_{3})^2 + \sqrt{x_{4}^2 + x_{5}^2}$, where, $x_{i}$ refers to the $i$'th feature, with $i = 1, \ldots, d$. 

\noindent {\bf Example 6:} This is a real-data example with $n=3449$ and $d=4$.
This dataset, taken from the paper~\cite{mekaroonreung2012estimating}, was downloaded from~\url{http://ampd.epa.gov/ampd/}.
The response was the amount of heat input with the covariates corresponding to the amounts of emissions of S02, NOx, C02 (in tons) and the NOX rate. There were some samples with 
missing values that were removed from the dataset. We took a logarithmic transformation of the covariates and observed (based on exploratory analysis) that the relationship between  the response and the individual covariates seemed to be modeled well via a convex fit. Both the response and features were centered and scaled as in the aforementioned 
instances.

\noindent {\bf Example 7:}
This example is a smaller subset of the dataset in Example 6. Here we took the first 1500 samples giving us: $n=1500, d= 5$.

\paragraph{Software Specifications:} All our computations were performed in~{\textsc{Matlab}}, 
 (R2014a (8.3.0.532) 64-bit (maci64)) which was interfaced with some of the matrix operations coded in \texttt{C} on a OS X 10.8.5 (12F45) operating system with a 3.4 GHz Intel Core i5 processor with 32 GB Ram,  processor speed 1600 MHz and DDR3 SDRAM.

For all the examples above, we applied Algorithm~1. We observed that the algorithm converged in all the instances, which was verified by checking that the conditions of optimality (see~\eqref{opt-conds-1}) were satisfied (approximately) up to algorithmic precision. The convergence speed was, however, found to be sensitive to the choice of $\rho$. We observed that after the data was standardized, i.e., features and responses set to have unit $\|\cdot\|_{2}$ norm, a value of $\rho$ of the order of $1/n$ performed quite well. Figure~\ref{fig:compute_times_fig1} shows some of the results for different values of $\rho \in \{{0.1}/n, {1}/{n}, {10}/{n} \}$; additional examples are presented in the appendix, see Figure~\ref{fig:compute_times_fig2} in Section~\ref{sec:AppB}.

Primal feasibility is measured by $\|\Gamma \|_{F}/n$, where,  $\Gamma = ((\gamma_{ij}))$ with $\gamma_{ij} = \eta^*_{ij}  - \left( \theta_{i}^* + \langle \Delta_{ij}, \B\xi_{j}^*\rangle \right.$ $\left. - \,\theta_{j}^* \right)$, as
defined in~\eqref{cond-11} and $\| \Gamma \|_{F}$ denotes the Frobenius norm of $\Gamma$. The gradient condition with respect to $\bt$ is measured by the $\|\cdot\|_{2}$-norm of the vector $(\B\theta^* - \M{Y}) - \M{D}^\top \text{vec}(\B\nu^*)$, as defined in~\eqref{cond-13}.
These are the two metrics that have been considered in Figure~\ref{fig:compute_times_fig1} and Figure~\ref{fig:compute_times_fig2}.
Figure~\ref{fig:compute_times_fig1} shows that there is no clear best choice of $\rho$, but there is however, one systematic pattern: a small choice of $\rho$ leads to faster changes in the objective value across iterations, but the primal feasibility goes to zero at a slower rate. Similarly, for a larger value of $\rho$ we observe that
the primal feasibility is (approximately) satisfied early on in the iterations, but it takes longer for the objective values to stabilize. As mentioned before, we typically found a choice of $\rho$ of the order of ${1}/{n}$ to work quite well in our experiments.

In addition to Algorithm~1, we have also considered Algorithm~2 which is theoretically guaranteed to converge. Numerical results showing the performance of Algorithm~1 versus Algorithm~2 is presented in Table~\ref{tab:algo1-algo3}. We have observed in our experiments that Algorithm~2 has similar empirical behavior as Algorithm~1.  In the implementation of Algorithm~2, we took a dynamically decreasing tolerance in primal feasibility and 
the gradient condition, following~\cite{aybat2012first}. For Algorithm~2, the first 500 (outer) iterations (as many dual variable updates) were taken to be identical to Algorithm~1; after which
each inner loop was executed till a tolerance of $\delta_{i}$  and  this tolerance was made tighter via the schedule $\delta_{i} \leftarrow \tau \delta_{i}$, with the starting value of $\delta_{i} = 10^{-1}$. 
Whenever the dual variable was updated, the  value of $\rho$ was increased by a factor $1/\tau$. We took $\tau = 0.9954$ and the algorithm was run for a maximum of 3000 outer iterations with an upper cap of 50 iterations for every inner loop.  
We find Algorithm~1 to be more appealing than Algorithm~2 because of its simplicity and its stable behavior for large values of $n$. However, if one indeed seeks a method with established convergence properties, we recommend the use 
of Algorithm~2. We consider these two methods to be close cousins of one another. 

%
%

\appendix

\section{Appendix}

\subsection{Proof of Lemma~\ref{lem:sm-approx0}}\label{proof-lem:sm-approx0}
\begin{proof}
The proof follows by observing that 
$\nabla \tilde{\phi}(\M x; \tau)  = \sum_{i=1}^{m} \hat{w}_{i}\M{a}_{i},$ where $\hat{\M{w}}$ is a maximizer of the optimization Problem~\eqref{add-conv-1}.
Now, observe that 
\begin{equation}\label{lab-11}
\| \nabla \tilde{\phi}(\M x; \tau) \|_{2} = \left \| \sum_{i=1}^{m} \hat{w}_{i}\M{a}_{i} \right\|_{2} \leq \sum_{i=1}^{m}  \hat{w}_{i} \| \M{a}_{i}\|_{2} \leq  \sum_{i=1}^{m}  \hat{w}_{i} L,
\end{equation}
where, above, we used the triangle inequality and the fact that $\| \M{a}_{i}\|_{2}\leq L$ for all $i$. The latter follows from the simple observation that 
each  $\M{a}_{i} =  \hat{\B\xi}_{i}$ and $\|  \hat{\B\xi}_{i} \|_{2} \leq L$ (since  $\hat{\B\xi}_{i}$'s are solutions to Problem~\eqref{eq:CvxLipLSE}).
We thus have, from~\eqref{lab-11}, that $\| \nabla \tilde{\phi}(\M x; \tau) \|_{2} \leq \sum_{i} \hat{w}_{i} L = L$.
\end{proof}

\subsection{Algorithm for Solving SOCP~\eqref{socp-ls-1}}\label{sec:append-socp-1}
Problem~\eqref{socp-ls-1} can be rewritten as the following optimization problem:
\begin{equation}\label{socp-2}
\mini_{\B\xi_{j}} \;\; \frac{1}{2} \| \B{A}_j \B{\xi}_{j} - \B{b}_j \|_2^2 \;\; s.t.  \;\; \| \B{\xi}_j \|^2_2 \leq L^2,
\end{equation}
where $\M{A}_j := [\B \Delta_{1j}; \B \Delta_{2j}; \ldots ; \B \Delta_{nj} ]$ is formed by stacking the vectors $
\B \Delta_{ij}$'s (for fixed $j$) into a matrix of size $n \times d$;
$\M{b}_{j}$ is obtained by vectorizing $\bar{\eta}_{ij}, i = 1, \ldots, n$.
Observe that the optimization Problem~\eqref{socp-2} can be equivalently written in the Lagrangian version:
\begin{equation} \label{Lipxi}
\mini_{\B\xi_{j}} \;\; \left( \frac{1}{2} \| \B{A}_j \B{\xi}_{j} - \B{b}_j \|_2^2 + \lambda'_j  \|\B \xi_j\|_2^2 \right)
\end{equation}
with Lagrangian parameter $\lambda'_j$. In fact, there is a choice of $\lambda'_{j}$ for which Problems~\eqref{Lipxi} and~\eqref{socp-2} are equivalent.
 We describe how to find  $\lambda'_j$ from $L$ by using a root-finding algorithm. 
Note that the solution to~\eqref{Lipxi} is given by
\begin{equation} \label{Lipxi2}
\B{\xi}^*_j (\lambda_{j})= \left (\B{A}^\top_j \B{A}_j + \lambda'_j I\right)^{-1}\B{A}^\top_j \B{b}_j.
\end{equation}
We will now simplify the above expression further. 
Consider the singular value decomposition of $\B A_j$ = $U_j \Gamma_j V^\top_j$ where $U_j$ and $V_j$ are $n\times d$ and $d\times d$ orthogonal matrices respectively and $\Gamma_j$ is a $d\times d$ diagonal matrix with diagonal entries $\gamma_i, i =1,...,d$. Then
\begin{equation} \label{Lipxi3}
\left (\B{A}^\top_j \B{A}_j + \lambda'_j I\right)^{-1} = V_j \Gamma'_j V^\top_j
\end{equation}
where $\Gamma'_j$ = $\text{diag}((\gamma_1^2+\lambda_j)^{-1},\ldots, (\gamma_d^2+\lambda_j)^{-1} )$. We have
\begin{eqnarray} \label{Lipxi4}
\B{\xi}^*_j (\lambda_{j})= (V_j \Gamma'_j V^\top_j)  V_j  \Gamma_j U^\top_j b_j  =     V_j \Gamma''_j (\lambda_{j}) U^\top_j \B b_j,
\end{eqnarray}
where $\Gamma''_j(\lambda_{j}) =\Gamma'_j \Gamma_j=\text{diag}(\gamma_1/(\gamma_1^2+\lambda_j), \ldots, (\gamma_d/(\gamma_d^2+\lambda_j))$,  and we used the fact that 
$V^\top_j V_j = I_d$, the $d\times d$ identity matrix. Note that $\| \B{\xi}_j (\lambda_{j})\|^2_2$ is given by
\begin{eqnarray} \label{Lipxi5}
\| \B{\xi}^*_j (\lambda_{j})\|^2_2 = \B b^\top_j U_j \Gamma'''_{j}(\lambda_{j})  U^\top_j \B b_j
\end{eqnarray}
where, $\Gamma'''_j(\lambda_{j}) = \text{diag}( \gamma^2_1/(\gamma_1^2+\lambda_j)^2, \ldots, \gamma^2_d/(\gamma_d^2+\lambda_j)^2 )$. We need to find a value of $\lambda_{j}$ for which $\| \B{\xi}^*_j (\lambda_{j})\|_{2} = L$, if such a value exists. Of course, the equality will not hold if 
$L > \| \B{\xi}^*_j (\lambda_{j})\|_{2}$ for all values of $\lambda$. Note that 
the largest value of $\| \B{\xi}^*_j(\lambda_{j}) \|_{2}$ is when $\lambda_{j}=0$, 
which corresponds to the unconstrained least squares solution $\| \B{\xi}^*_j(0) \|_{2}$.
Thus given a value of $L$ we first need to 
check if $L$ is larger than  $\| \B{\xi}^*_j(0) \|_{2}$. If yes, then the solution to Problem~\eqref{socp-2} will be $\B{\xi}^*_j(0)$.
Otherwise, there exists a value of $\lambda_{j} >0$ for which the equality $\| \B{\xi}^*_j(\lambda_{j})\|_{2}=L$ holds --- this corresponds to the case where the Lipschitz regularization is effective.
For finding the value of $\lambda_{j} >0 $ that corresponds to the given $L$, we employ a standard Newton-Raphson type root-finding method (see e.g.,~\cite{nesterov2004introductorynew})
to search for $\lambda_{j}$ that satisfies the equality
$\B b^\top_j U_j \Gamma'''_j(\lambda_{j})  U^\top_j \B b_j = L^2$, up to numerical precision. Once the $\lambda^*_j$ that satisfies $\| \B \xi^*_j(\lambda^*_{j})\|_{2} = L$ is obtained,  the corresponding 
 $\B \xi^*_j(\lambda^*_{j})$ gives us a solution to Problem~\eqref{socp-ls-1}.

Note that the main computational cost in this problem lies in doing an SVD of $\M{A}_{j}$ with cost $O(nd^2)$ (assuming $n \gg d$). However, this can be pre-computed before running the ADMM algorithm. The $\B{b}_{j}$'s need to be updated after every ADMM iteration, but once the vectors $U_{j}^\top\B{b}_{j}$'s are computed with cost $O(nd)$, the different Newton-Raphson iterations cost $O(d)$ for computing the function values and gradients. Thus, the overall cost of the algorithm described above is $O(nd)$, in addition to the SVD computations which can be done off-line. Note that there are $n$ different vectors $\B\xi_{j}$'s that need to be updated leading to an overall cost of $O(n^2d)$ for the update in all the $\B\xi_{j}$'s.

\subsection{A coordinate descent algorithm for Problem~\eqref{nng-ls-1}}\label{appendix-cd-1}
Problem~\eqref{nng-ls-1} is a special instance of the following non-negative LS problem:
\begin{equation}
\mini_{\M{u} \in \Re^d} \;\; \M{u}^\top Q \M{u} + \langle \M{a}, \M{u} \rangle \;\;\; s.t. \;\;\; \M{u} \geq 0,   
\end{equation}
which involves the minimization of a convex quadratic $Q \succeq \M{0}$ function with separable constraints.  Due to the smoothness of the convex objective function and separability of the constraints, one-at-a-time coordinate descent can be employed for the problem; see e.g.,~\cite{tseng-92},~\cite{FHT2007}. 
\medskip

\begin{algorithm}
 \caption{Coordinate descent for Problem~\eqref{nng-ls-1}}\label{cd-algo-1}
\begin{enumerate}
\item Start with $\M{u}^1 \geq 0$. For $ m \geq 1$ do the following 
 until
$\| \M{u}^{m+1} - \M{u}^m \|_{2}  \leq \text{TOL} \| \M{u}^m \|_{2}$ for some
predefined tolerance ``TOL''.

\item Assign $ \M{u} \leftarrow \M{u}^m $ and for $k \in \{1, \ldots, d\}$ do the following
\begin{enumerate}
\item Fix $u_{\ell}$ for $\ell \neq k$ and update
$$ u_{k} = \argmin_{ u \geq 0 }  \;\; \left (q_{kk}u^2 + \tilde{a}_{k}  u\right)\; = \max \left\{ -\frac{\tilde{a}_{k}}{2 q_{kk}} , 0  \right\}, $$
where $\tilde{a}_{k} = 2 \sum_{\ell \neq k} q_{k \ell} u_{\ell} + a_{k}$ and $Q=((q_{ij}))$.
\end{enumerate}

\item Assign $\M{u}^{m +1 } \leftarrow \M{u}$ and go to Step 2.

\end{enumerate}
\end{algorithm}

Note that the vector $\tilde{\M{a}} =(\tilde{a}_{1}, \ldots, \tilde{a}_{d})$ is defined as $\tilde{\M{a}}:=Q\M{u} + \M{a}$ and once the $k$'th coordinate $u_{k}$ is updated by $\delta_{k}$ (say),  
then $\tilde{\M{a}} \leftarrow \tilde{\M{a}} + Q[,k]\delta_{k}$ where, $Q[,k]$ denotes the $k$'th column of the matrix $Q$ and the whole update in $\tilde{\M{a}}$ can be performed in $O(d)$ operations. Thus for one full cycle over the $d$ coordinates the total cost is $O(d^2)$. Note that in the process of updating the coordinates $u_{k}$ many of the coordinates that were at zero stay at zero, thus no updating is required for that coordinate --- this leads to sparse updating rules in $u_{k}$ and often leads to significantly improved  computational performance. More precisely, if out of the $d$ coordinates a few of the coordinates need to be updated the total cost of performing Step 2 reduces to $O(d)$. The interested reader can also see~\cite{FHT2007} for related computational tricks employed in $\ell_{1}$-regularized problems.

Note that Step 1 of the algorithm requires a starting vector $\M{u}^1$. Since the above algorithm is used as a part of the ADMM algorithm, the current solution of Problem~\eqref{nng-ls-1} can be used as a warm-start for the above algorithm. This often leads to performing fewer cycles across the $d$ coordinates.

\subsection{Proof of Lemma~\ref{lem:sm-approx1}}\label{proof:lem:sm-approx1}
\begin{proof}
The proof follows by observing that 
$\nabla \tilde{\phi}(\M x; \tau)  = \sum_{i=1}^{m} \hat{w}_{i}\M{a}_{i},$ where $\hat{\M{w}}$ is a maximizer of the optimization Problem~\eqref{add-conv-1}.
Since $\hat{\M{w}} \in \Delta_{m}$, every coordinate of $\nabla \tilde{\phi}(\M x; \tau)$ is a convex combination of the coordinates of $\M{a}_{1},\ldots, \M{a}_{m}$.
Note that by~\eqref{aibi-choice}, for every $i=1, \ldots, m$, we have $\M{a}_{i} = \hat{\B\xi}_{i} \geq 0$ (since it is a solution to Problem~\eqref{conv-reg1}). Thus, for every coordinate 
$k$,  we have $\nabla_{k} \tilde{\phi}(\M x; \tau) \geq 0$. This completes the proof of the lemma.
\end{proof}

\subsection{Proof of Theorem~\ref{thm:RateOfConv}}\label{proof:RateOfConv}
\begin{proof}
The theorem follows from known metric entropy results on the class of uniformly bounded convex functions that are uniformly Lipschitz in conjunction with known results on the rates of convergence of LSEs; see e.g.,~\citet[Theorem 9.1]{vdG00}. We give the details below. 

The notion of covering numbers will be useful. For $\epsilon > 0$ and a subset $S$ of functions, the $\epsilon$-covering number of $S$ under the metric $\ell$, denoted by $N(S, \epsilon;\ell)$, is defined as the smallest number of closed balls of radius $\epsilon$ (under the metric $\ell$) whose union contains $S$.

Fix any $B>0$ and $L> L_0$. We define the class of uniformly bounded convex functions that are uniformly Lipschitz as
\begin{equation}\label{eq:C_L}
\C_{L,B} := \{\psi \in \C_L: \|\psi\|_\mathfrak{X} \le B\},
\end{equation} 
where $\|\psi\|_\mathfrak{X} := \sup_{x \in \mathfrak{X}} |\psi(x)|$.
Using Theorem 3.2 of \citet{GS13} (also see \citet{Bronshtein76}) we know that   
\begin{eqnarray}\label{difflip.eq}
   \log N \left(\C_{L, B}, \epsilon; \ell_{\infty} \right)  \leq c \left(\frac{B + d L}{\epsilon}  \right)^{d/2},
  \end{eqnarray}
for all $0 < \epsilon \leq \epsilon_0 (B + d L)$, where $\epsilon_0 >0$ is a fixed constant and $\ell_\infty$ is the uniform metric. Let $\hat \phi_{n,L,B} $ denote a LSE of $\phi$ in the class $\C_{L,B}$. Routine calculations and Theorem 9.1 of \citet{vdG00} now yields
\begin{equation}\label{eq:L_2RateBdLip}
	\frac{1}{n} \sum_{i=1}^n (\hat \phi_{n,L,B}(\B X_i) - \phi(\B X_i))^2 = O_\pr(r_n),
\end{equation}
where $r_n$ is defined as in~\eqref{eq:Rates}. 

Define the event $A_n := \{ \max_{i=1,\ldots,n}|\hat{\phi}_{n,L}(\B X_i)|\le B_0 \}$. Next we show that there exists $B_0>0$ such that 
\begin{equation}\label{eq:A_n}
\pr(A_n) \rightarrow 1, \qquad \mbox{as } n \rightarrow \infty.
\end{equation} 
From the characterization of the projection on the closed convex set $\C_L$, we know that $$\sum_{i=1}^n (Y_i - \hat \phi_{n,L}(\B X_i)) (\gamma(\B X_i) - \hat \phi_{n,L}(\B X_i)) \le 0,$$ for all $\B \gamma \in \C_L$. Letting $\M{e} \equiv 1$ denote the constant 1 convex function, note that for any $c \in \Re$, $c  \M{e} \in \C_L$. Hence simple algebra yields $\sum_{i=1}^n(Y_i - \hat \phi_{n,L}(\B X_i)) \M{e}(\B X_i)= 0,$ i.e., $\sum_{i=1}^n Y_i = \sum_{i=1}^n \hat \phi_{n,L}(\B X_i)$. Now, letting $\bar Y = \sum_{i=1}^n Y_i/n$, for any $\M x \in \mathfrak{X}$,
\begin{eqnarray*}
	|\hat \phi_{n,L}(\M x)| & \le & |\hat \phi_{n,L}(\M x) - \bar Y| + |\bar Y|  \;\; =\;\; \left|\hat \phi_{n,L}(\M x) - \frac{1}{n} \sum_{j=1}^n\hat \phi_{n,L}(\M X_j) \right| + |\bar Y| \\
	& \le & \frac{1}{n} \sum_{j=1}^n  \left|\hat \phi_{n,L}(\M x) - \hat\phi_{n,L}(\M X_j) \right| + |\bar Y| \\
	& \le & \frac{L}{n} \sum_{j=1}^n  \|\M x - \B X_j \|_2 + \|\phi\|_\mathfrak{X} + |\bar \epsilon| \\
	& \le & \sqrt{d} L + \kappa +1 =: B_0,
\end{eqnarray*}
a.s.~for large enough $n$, where we have used the fact that $\|\M x - \M X_j \|_2 \le \sqrt{d}$, $\|\phi\|_\mathfrak{X} < \kappa$ for some $\kappa >0$, and that $\bar \epsilon := \sum_{i=1}^n \epsilon_i/n \rightarrow 0$ a.s. As $\C_{L,B_0} \subset \C_L$, we trivially have $$\sum_{i=1}^n (\hat \phi_{n,L}(\M X_i) - Y_i)^2 \le \sum_{i=1}^n (\hat \phi_{n,L,B_0}(\B X_i) - Y_i)^2.$$ If $A_n$ happens, $\hat \phi_{n,L} \in \C_{L,B_0}$, and thus, $$\sum_{i=1}^n (\hat \phi_{n,L}(\M X_i) - Y_i)^2 \ge \sum_{i=1}^n (\hat \phi_{n,L,B_0}(\M X_i) - Y_i)^2.$$ From the last two inequalities and the uniqueness of the projections it follows that if $A_n$ occurs, then $\hat \phi_{n,L} = \hat \phi_{n,L,B_0}$ at the data points. Now using \eqref{eq:A_n}, \eqref{eq:Rates} immediately follows from \eqref{eq:L_2RateBdLip}.
\end{proof}

\section{Additional Computational Results} \label{sec:add-comp-results1}\label{sec:AppB}

\begin{figure}[h!]
\centering
\resizebox{\textwidth}{!}{\begin{tabular}{ l c c  cc}
&\scriptsize \sf Example 3 &\scriptsize \sf Example 4 &\scriptsize \sf Example 5 & \scriptsize \sf Example 11 \\
\rotatebox{90}{\sf {\scriptsize{~~~~~~~~$\log_{10}$(Primal Feasibility)}}}&
\includegraphics[width=0.24\textwidth,height=0.22\textheight,  trim =1.0cm 2.5cm 1.cm 1.5cm, clip = true ]{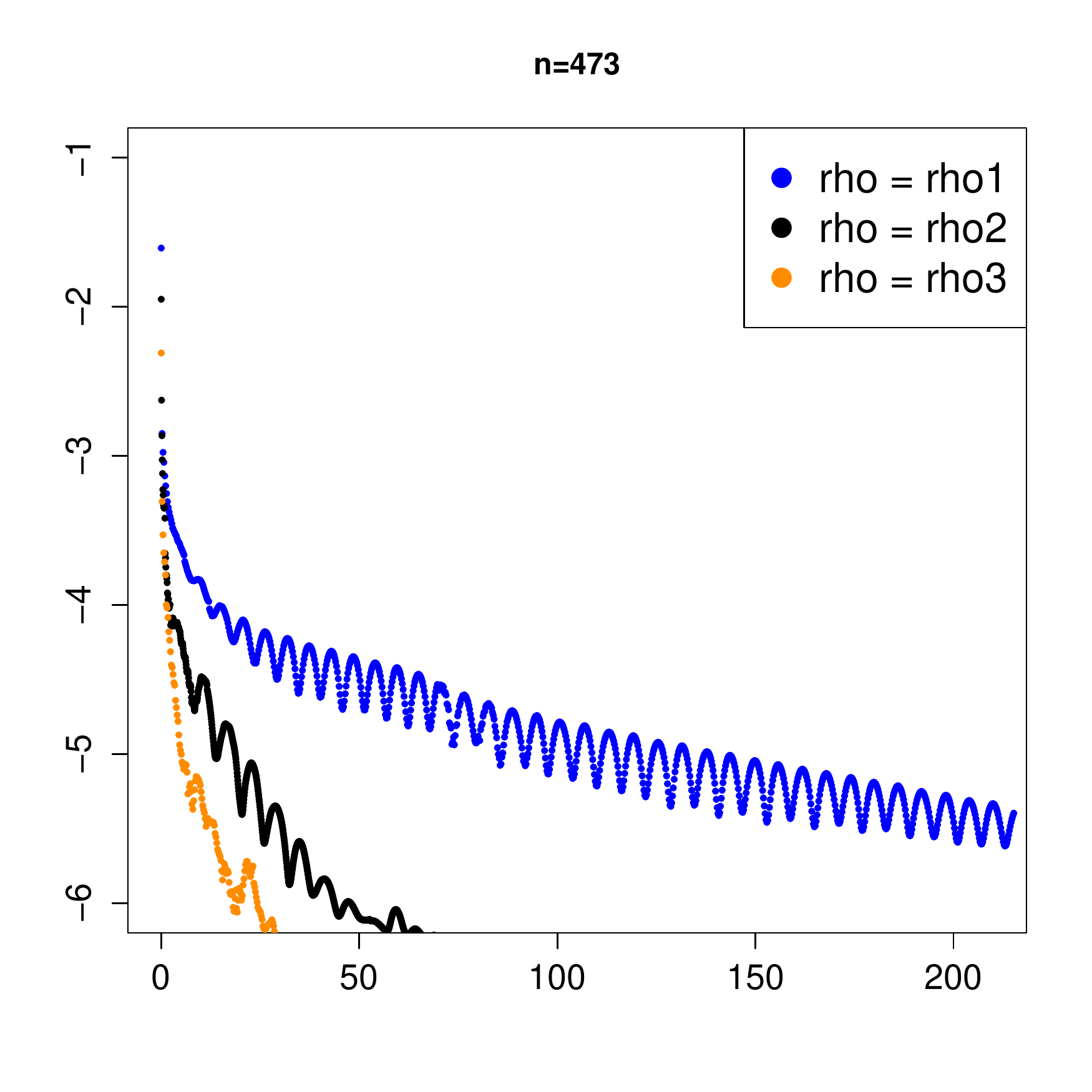}&
\includegraphics[width=0.24\textwidth,height=0.22\textheight,  trim =2.0cm 2.5cm 1.cm 1.5cm, clip = true ]{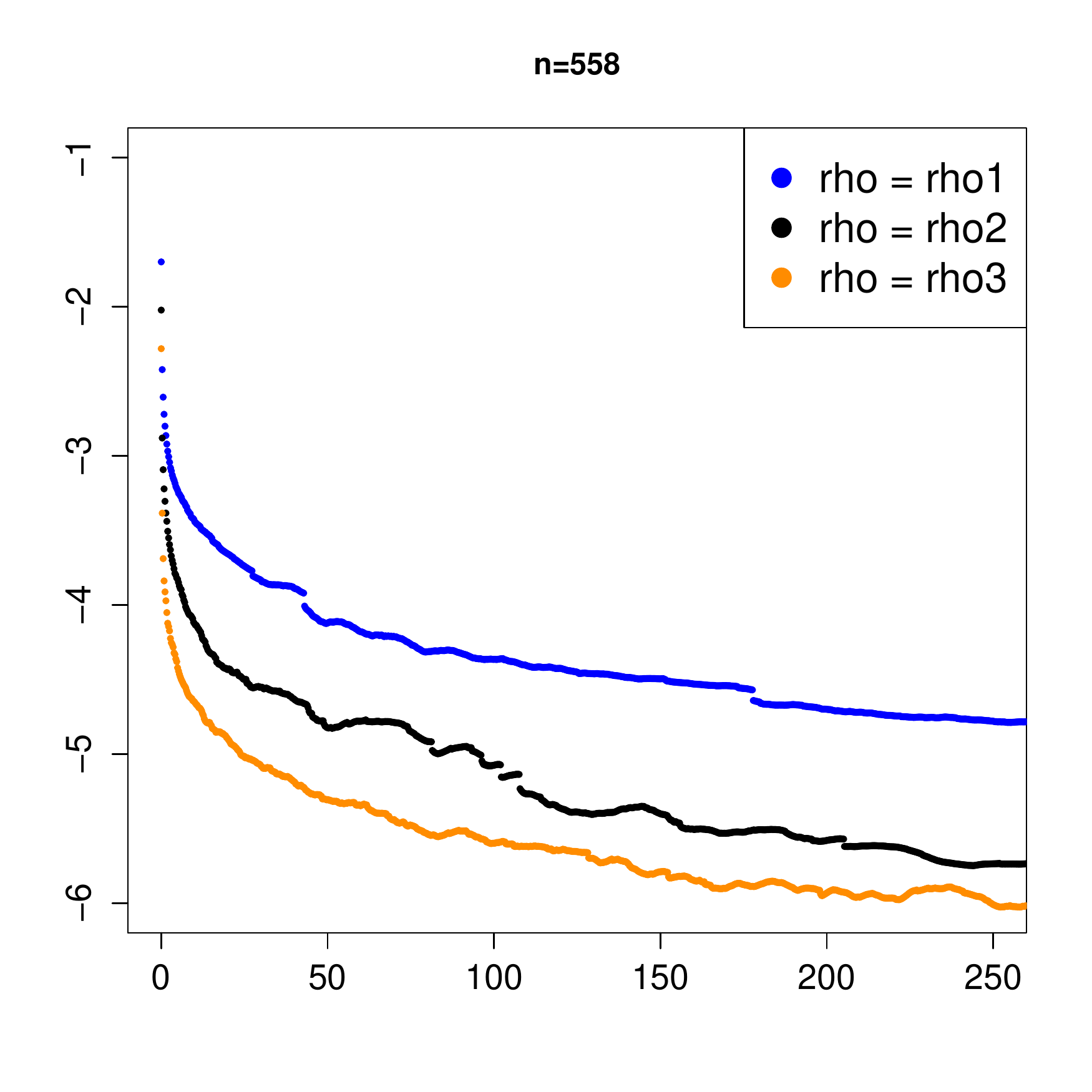}&
\includegraphics[width=0.24\textwidth,height=0.22\textheight,  trim =2.0cm 2.5cm 1.cm 1.5cm, clip = true ]{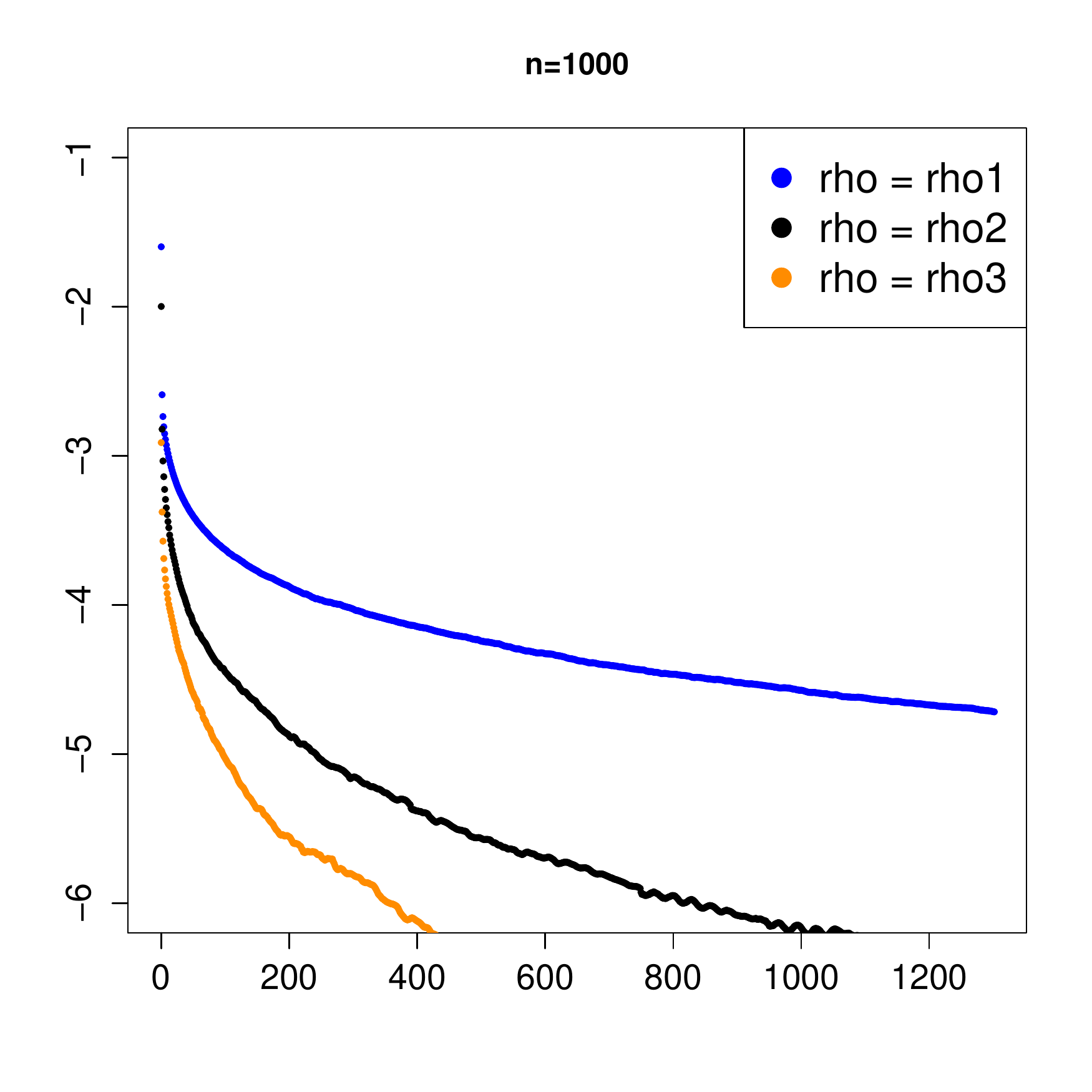} &
\includegraphics[width=0.24\textwidth,height=0.22\textheight,  trim =2.0cm 2.5cm 1.cm  1.5cm, clip = true ]{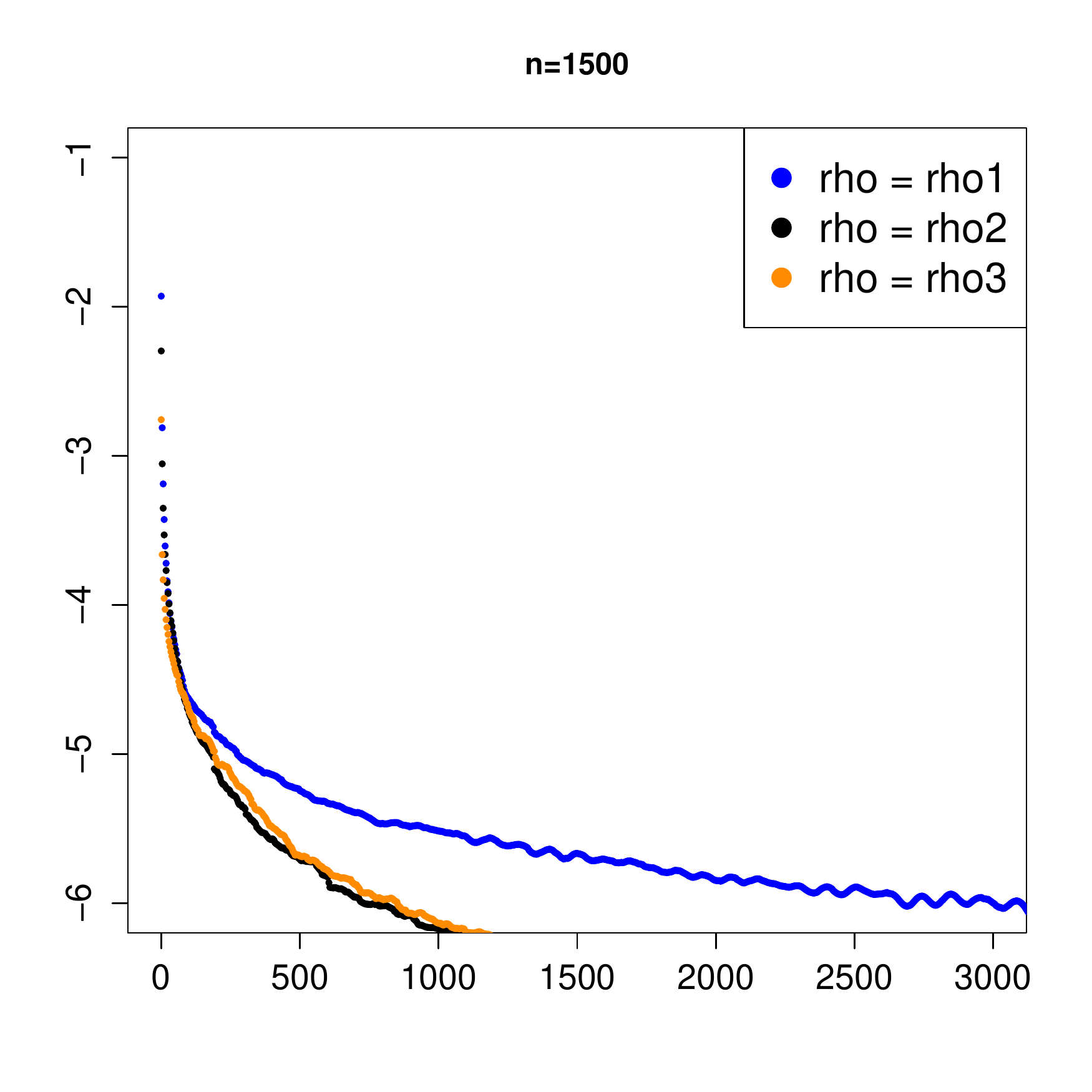} \\
\rotatebox{90}{\sf {\scriptsize{~~~~~~~~~~~~~$\log_{10}$(Gradient)}}}&
\includegraphics[width=0.24\textwidth,height=0.22\textheight,  trim =1.0cm 1.5cm 1.cm 1.5cm, clip = true ]{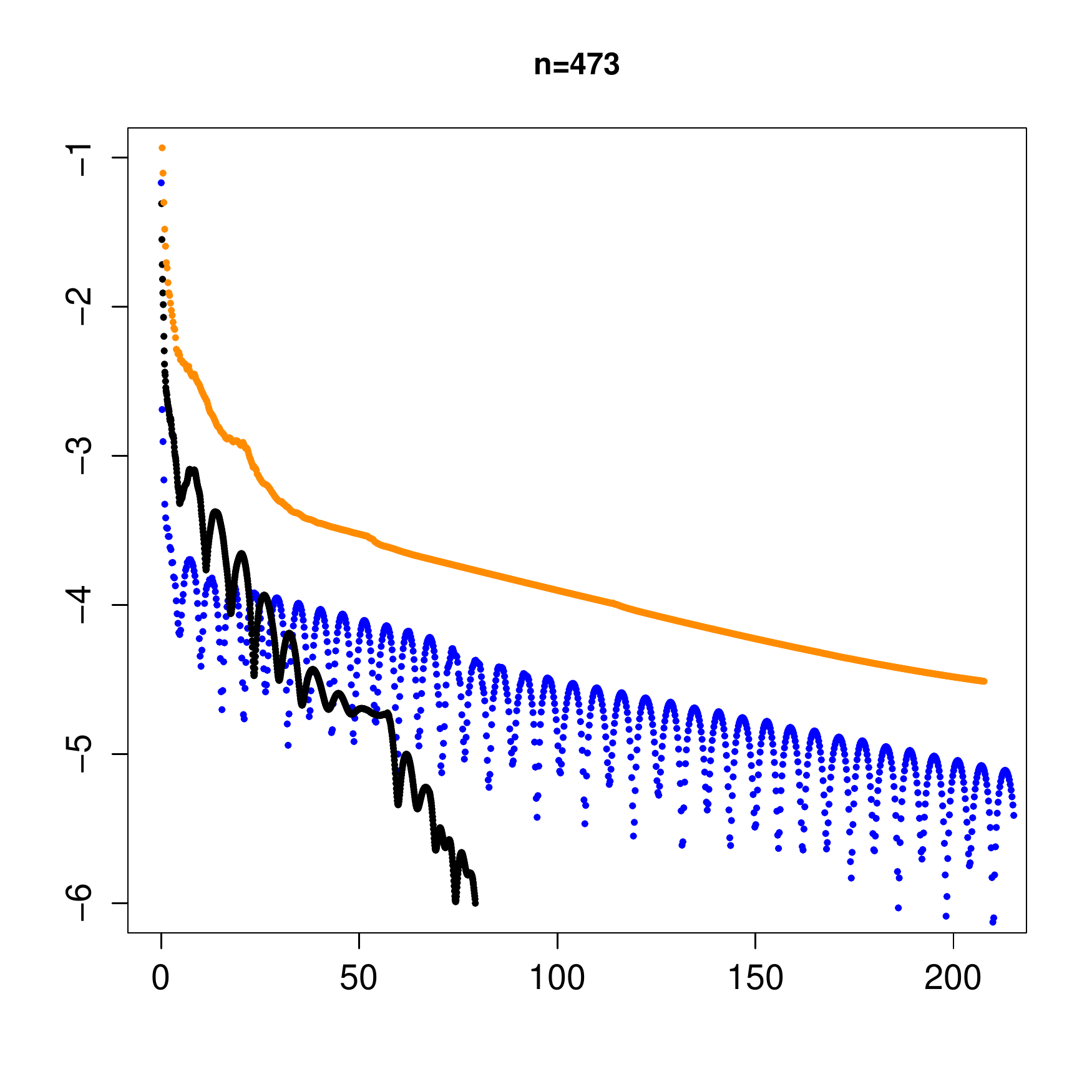}&
\includegraphics[width=0.24\textwidth,height=0.22\textheight,  trim =2.0cm 1.5cm 1.cm 1.5cm, clip = true ]{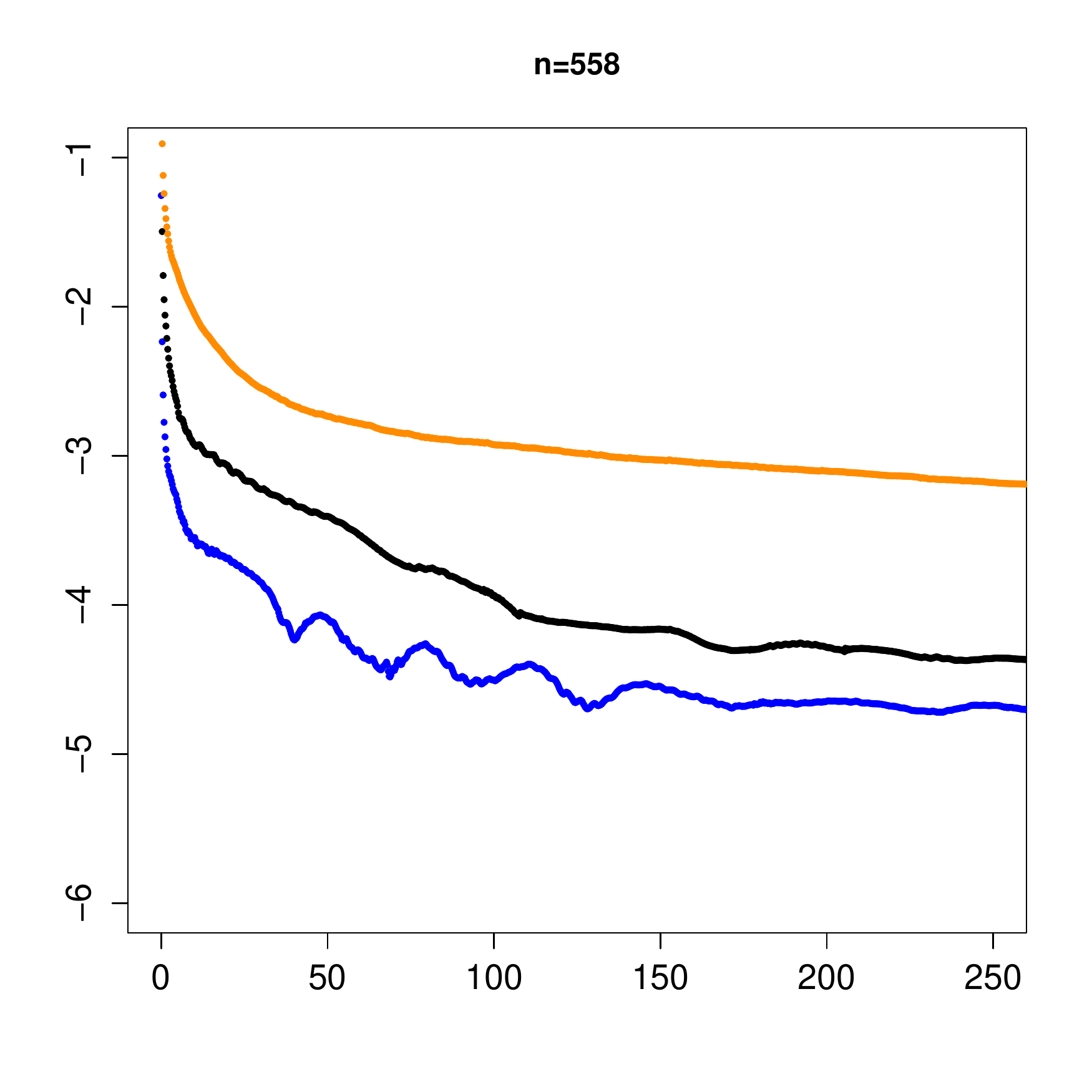}&
\includegraphics[width=0.24\textwidth,height=0.22\textheight,  trim =2.0cm 1.5cm 1.cm 1.5cm, clip = true ]{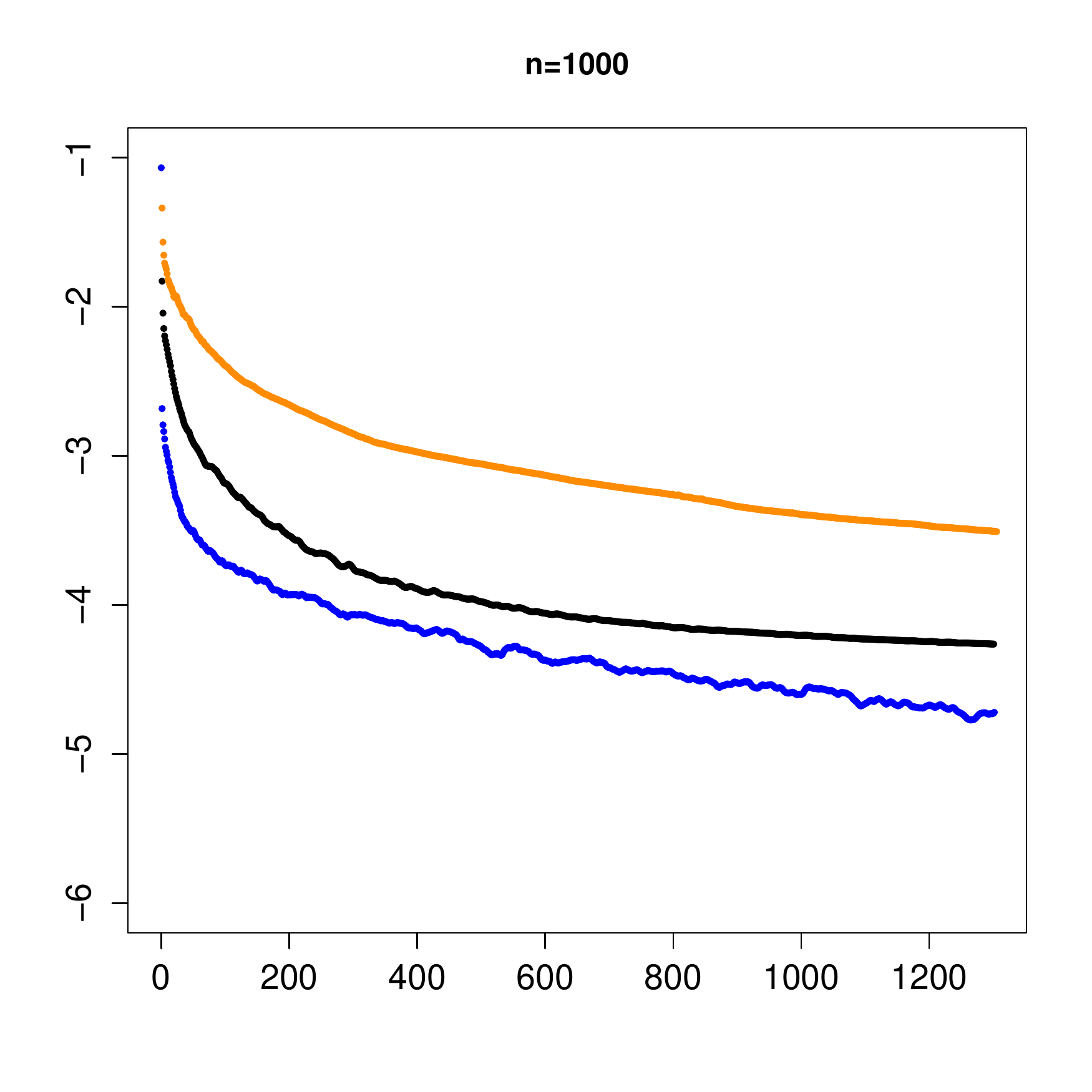}&
\includegraphics[width=0.24\textwidth,height=0.22\textheight,  trim =2.0cm 1.5cm 1.cm 1.5cm, clip = true ]{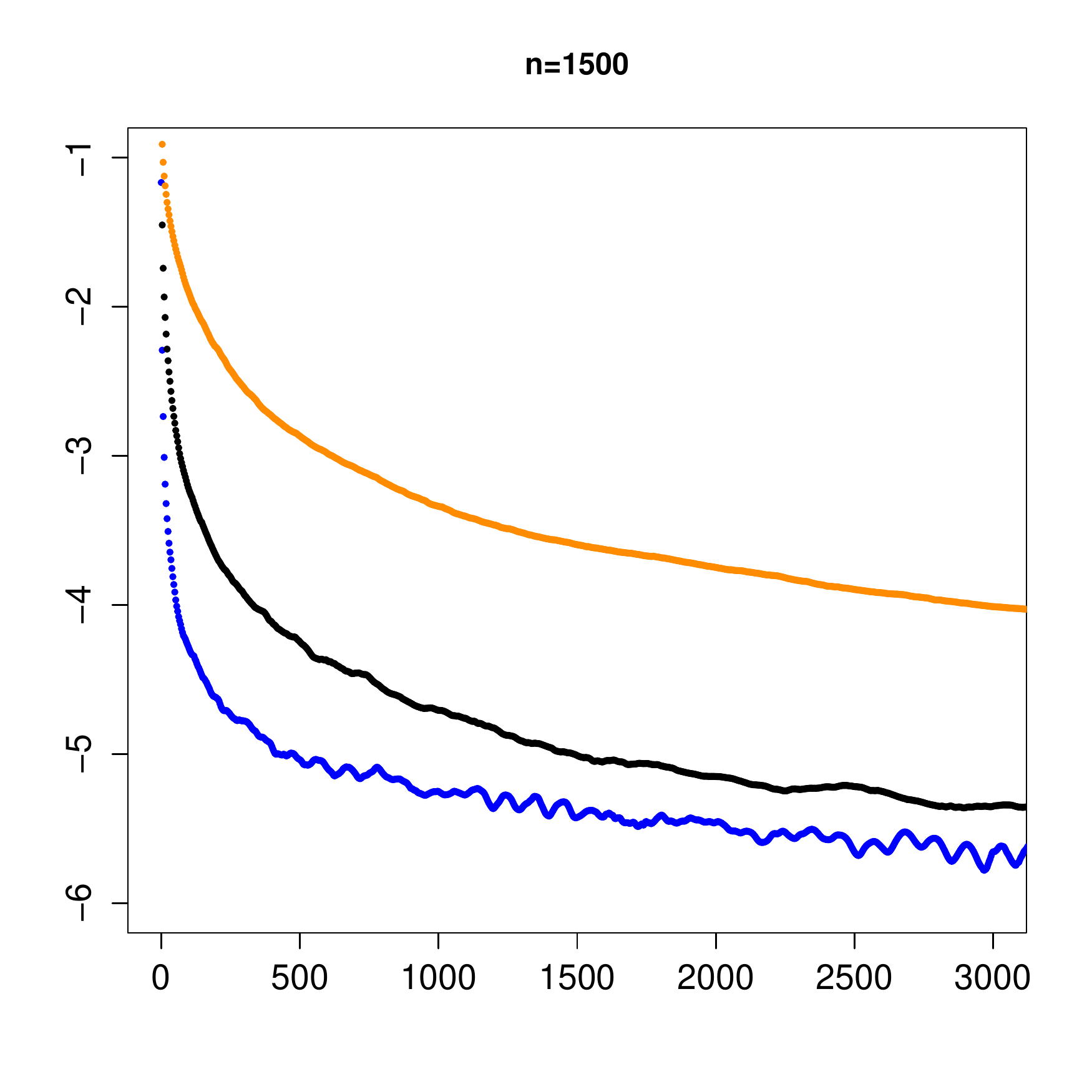} \\
&\scriptsize \sf Time (secs) &\scriptsize \sf Time (secs) &\scriptsize \sf Time (secs) & \scriptsize \sf Time (secs) \\
\end{tabular}}
\caption{Figure showing the convergence characteristics of Algorithm~1 for different examples, as described in the text. The legends have the same meanings as in Figure~\ref{fig:compute_times_fig1}.}\label{fig:compute_times_fig2}
\end{figure}

\begin{spacing}{1.05}
{
\bibliographystyle{Chicago}
\bibliography{Referencias}}
\end{spacing}
\end{document}